\documentclass[review]{elsarticle}

\usepackage{amsmath}
\usepackage{pgfplots}
\usepackage[hidelinks]{hyperref}

\usepackage{amssymb}
\usepackage{graphicx}
\usepackage{grffile}
\usepackage{epstopdf}
\DeclareGraphicsExtensions{.eps}
\usepackage{epsfig} 
\usepackage{subcaption}

\usepackage{longtable} 
\usepackage{pdflscape} 
\usepackage{rotating} 
\usepackage{booktabs}
\usepackage{multirow}
\usepackage{tabularx}
\usepackage{xcolor}

\usepackage[absolute,overlay]{textpos}

\usepackage[parfill]{parskip}
\usepackage[german]{varioref}
\usepackage{url}

\usepackage{dcolumn}
\usepackage{tikz}
\usetikzlibrary{shapes,arrows}
\usetikzlibrary{intersections}
\usepackage{upgreek}

\usepackage[margin=1.44in]{geometry}

\usepackage{hypernat}

\usepackage{titlesec}
\titleformat{\paragraph}
{\normalfont\normalsize\itshape}{\theparagraph}{1em}{}
\titlespacing*{\paragraph}
{0pt}{3.25ex plus 1ex minus .2ex}{1.5ex plus .2ex}

\usepackage{algorithm}
\usepackage{algpseudocode}
\usetikzlibrary{shapes,arrows,arrows.meta,positioning,
	fit,backgrounds,calc}
	
\newcolumntype{f}{>{$}l<{$}}
\newcolumntype{n}{l}
\newcolumntype{N}{>{\scriptsize}l}
\newcolumntype{v}[1]{>{\raggedright\hspace{0pt}}p{#1}}
\newcolumntype{V}[1]{>{\scriptsize\raggedright\hspace{0pt}}p{#1}}
\newcolumntype{C}[1]{>{\scriptsize\centering\hspace{0pt}}p{#1}}
\newcolumntype{B}[1]{>{\boldmath\DC@{.}{,}{#1}}l<{\DC@end}}
\newcolumntype{d}[1]{>{\DC@{.}{,}{#1}}l<{\DC@end}}
\newcolumntype{R}[1]{%
	>{\begin{turn}{90}\begin{minipage}{#1}\scriptsize\raggedright\hspace{0pt}}l%
			<{\end{minipage}\end{turn}}%
}
\newcolumntype{x}{>{\scriptsize\raggedright\hspace{0pt}}X}

\journal{Composite Structures}

\begin{document}


\begin{frontmatter}


\title{\large Bridging nanoparticle morphology and viscoelastic behavior in epoxy nanocomposites: A coarse-grained simulation-informed constitutive model}

\author[mymainaddress]{Atiyeh Hente\corref{mycorrespondingauthor}}
\cortext[mycorrespondingauthor]{Corresponding author}
\ead{a.hente@isd.uni-hannover.de}
\author[mysecondryaddress]{Shadab Zakavati}
\author[mysecondryaddress,mythirdaddress]{Behrouz Arash}
\author[myfourthaddress]{Maximilian Jux}
\author[mymainaddress]{Raimund Rolfes}
\address[mymainaddress]{Institute of Structural Analysis, Leibniz University Hannover, Appelstra{\ss}e 9A, 30167 Hannover, Germany}
\address[mysecondryaddress]{Department of Mechanical, Electrical, and Chemical Engineering, Oslo Metropolitan University, Pilestredet 35, 0166 Oslo, Norway}
\address[mythirdaddress]{Green Energy Lab, Department of Mechanical, Electrical and Chemical Engineering, OsloMet – Oslo Metropolitan University, Oslo, Norway}
\address[myfourthaddress]{Institute of Composite Structures and Adaptive Systems, DLR (German Aerospace Center), Lilienthalplatz 7, 38108 Brunswick, Niedersachsen, Germany}

\begin{abstract}

Accurate prediction of the material behavior of polymer nanocomposites under various thermomechanical loading conditions is increasingly demanded for engineering applications. This study proposes an integrated framework combining coarse-grained (CG) molecular simulations and experimental testing to develop predictive constitutive models for nanoparticle/epoxy nanocomposites. The key contribution of this work lies in characterizing the influence of nanoparticle content and agglomerate size on the rate- and temperature-dependent behavior of nanocomposites, enabled by large-scale CG simulations. The proposed framework successfully captures the material response, including nonlinear hyperelasticity, softening behavior, and rate- and temperature-dependent properties, across a broad range of strain rates, temperatures, and nanoparticle sizes and weight fractions. The predictive capability of the CG simulation-informed constitutive model is validated using additional experimental data that were not included in the parameter identification process. By reducing reliance on extensive experimental testing while maintaining high accuracy, this simulation-driven approach offers an efficient pathway for developing robust, predictive constitutive models for designing and optimizing advanced nanocomposites.

\end{abstract}

\end{frontmatter}



\section{Introduction}


Polymer nanocomposites (PNCs) consist of a polymer matrix reinforced with fillers possessing at least one nanoscale dimension \cite{zaferani2018introduction} and offer an attractive combination of low density, enhanced strength, ductility, and thermal stability \cite{fu2008effects}, making them well-suited for demanding applications in aerospace structures, wind turbine rotor blades, and lightweight automotive components. 
Among various nano-fillers, boehmite nanoparticles (BNPs) are particularly effective for reinforcing epoxy matrices~\cite{jux2017effects,unger2020effect}. Even at low loadings, BNPs can significantly enhance tensile modulus, strength, and fracture toughness compared with neat epoxies~\cite{fankhanel2016mechanical,kausar2021review}. 

These improvements arise from the high surface area of BNPs, although agglomeration during processing can reduce their effectiveness \cite{jux2017effects,zare2017accounting}. The mechanical performance of BNP/epoxy nanocomposites is therefore governed by nanoparticle dispersion, agglomerate size, distribution, and filler weight fraction \cite{uddin2010improved,sapiai2015effect}.
Recent modeling efforts have further advanced understanding of BNP/epoxy behavior. Arash et al. \cite{arash2024cyclic} developed a phase-field model to study the cyclic viscoelastic–viscoplastic response under temperature and moisture variations. Hente et al. \cite{hente2021optimization} introduced a coarse-grained (CG) model to capture nanoscale interactions in agglomerated nanocomposites, while Hente et al. \cite{hente2025enhancement} advanced this approach using machine-learning-assisted coarse-graining to predict fracture enhancements. These studies collectively highlight that the mechanical response of BNP/epoxy nanocomposites depends on nanoscale morphology and environmental conditions, emphasizing the limitations of relying solely on experiments for characterization.

Due to the complex interactions between nanoparticles and polymer matrices, fully characterizing PNCs through experiments alone is challenging. All-atom simulations provide nanoscale insights \cite{li2013investigation,jang2014interfacial,li2016study,unger2019molecular, chen2025molecular}, but their high computational cost and limited length and time scales restrict studies of nanoparticle aggregation and distribution effects on viscoelastic behavior. CG simulations address these limitations by mapping groups of atoms to superatoms, extending time and length scales while retaining essential physical interactions \cite{arash2015mechanical,mousavi2016coarse}. CG simulations have also been employed to investigate the mechanical behavior of fiber-reinforced polymer composites, including tensile fracture of carbon nanotube reinforced polymers \cite{arash2015tensile} and modulus transition and debonding behavior at fiber-resin interfaces \cite{li2025coarse}.
Recent methodological advances have further enhanced the predictive power of CG simulations. Machine learning has been integrated to accelerate predictions and capture complex polymer behaviors \cite{ricci2023integrating,zhang2024chemically,fu2022simulate}, while Bayesian optimization enables rapid and systematic tuning of CG force fields and molecular topologies \cite{weeratunge2023bayesian, ray2025refining}. Neural network-based potentials provide an accurate representation of polymer dynamics and mechanical responses \cite{ivanov2023coarse,wilson2023anisotropic}. 

Building on these molecular- and CG-level insights, physically based constitutive models~\cite{arruda1993three,kontou2006viscoplastic,drozdov2002effect} have been developed to translate nanoscale mechanisms into predictive macroscale descriptions of PNC behavior under varying temperatures, strain rates, and network topologies.
These models are particularly valuable since their parameters are related to fundamental mechanisms and physical quantities at small scales, thereby linking macroscopic observations with molecular phenomena for accurate descriptions of material behavior. 
Arash et al. \cite{arash2019viscoelastic} proposed a viscoelastic damage model for BNP/epoxy nanocomposites in which atomistic simulations combined with experimental data were used for parameter identification. The model successfully captured key features of the stress–strain response of PNCs, such as nonlinear hyperelastic, time-dependent, and softening behavior. This approach was later extended to short-fiber reinforcements \cite{arash2021viscoelastic} and further developed into a finite deformation gradient-enhanced damage model for BNP/epoxy nanocomposites \cite{arash2021finite}. More recently, new contributions have advanced constitutive modeling, including multiscale frameworks \cite{yin2023multiscale}, viscoelastic–viscoplastic formulations \cite{yazdanparast20243d}, hygrothermal models \cite{xu2024development}, and machine learning- or physics-informed approaches \cite{bahtiri2023machine,bahtiri2024thermodynamically,upadhyay2024physics}. 
These studies demonstrate the growing capability of constitutive models to capture the nonlinear, time-dependent, and environment-sensitive response of PNCs. 

Yet, a key challenge remains in linking nanoscale morphology to macroscopic laws. The gap is often bridged through strain amplification models. Strain amplification–based constitutive models relate a macroscopically imposed strain to the average strain in a nanoparticle-filled epoxy matrix, following concepts introduced by Guth \cite{guth1938hydrodynamical, guth1945theory}. They provide simple yet effective relationships, such as between the modulus of the filled epoxy and filler volume fraction, though mainly under restrictive conditions (e.g., spherical fillers, low volume fractions, poorly reinforcing particles). Current models \cite{arash2019viscoelastic, arash2021viscoelastic, arash2021finite, unger2020effect} accurately predict behavior for well-dispersed nanoparticles. However, no framework fully integrates CG simulation insights across agglomeration scales with constitutive modeling that captures rate-, temperature-, and size-dependent effects. Recent hybrid approaches combining multiscale, morphology-informed constitutive frameworks with data-driven techniques \cite{yin2023multiscale, yazdanparast20243d} point toward solutions. Nevertheless, a comprehensive model simultaneously considering nanoparticle size, distribution, and rate- and temperature-dependent behavior is still lacking. Building on these insights, the present study combines CG simulations and experimental data to develop a predictive constitutive model for the complex, multiscale behavior of BNP/epoxy nanocomposites.

The present study proposes a combined framework of CG simulations and experimental testing to develop a constitutive model for BNP/epoxy nanocomposites. This approach enables characterization of the effects of BNP weight fraction and agglomerate size on the nanocomposites’ nonlinear viscoelastic behavior across a wide range of temperatures, strain rates, and nanoparticle loadings. CG simulations of epoxy matrices containing both well-dispersed and agglomerated BNPs provide lower and upper bounds for the stress–strain response of the nanocomposites. To account for temperature-dependent effects on rate-dependent and rate-independent responses, the Argon viscoelastic model and a modified Kitagawa model are employed. Calibration of these models demonstrates that the proposed constitutive framework accurately predicts the mechanical response of BNP/epoxy nanocomposites over a broad spectrum of strain rates, temperatures, nanoparticle sizes, and distributions. Overall, the simulation-based approach not only reduces the number of experiments required for model calibration but also shows good predictive capability. 

The present study advances the state of the art in the following key aspects. First, the Argon viscoelastic model parameters and the strain amplification factor are expressed as closed-form functions of both BNP weight fraction and agglomerate size, derived directly from large-scale CG simulations without requiring additional atomistic calibration. Second, the proposed framework simultaneously captures rate-, temperature-, and size-dependent effects of nanoparticle agglomeration within a single constitutive model, which to the authors' knowledge has not been reported before. Third, the predictive capability of the model is validated against experimental data that were entirely excluded from the parameter identification process, providing an independent and stringent assessment of the model's generalization capability.
The outline of the paper is as follows. Section~\ref{section:Constitutive_model} presents the constitutive model. The material system and coarse-grained modeling approach are described in Sections~\ref{section:Material} and ~\ref{section:cg}, respectively. The experimental program is outlined in Section~\ref{section:ExP_explain}. Results, parameter identification, and experimental validation are discussed in Section~\ref{section:Results}, followed by concluding remarks in Section~\ref{section:conclusions}.


\section{Constitutive model for nanoparticle reinforced epoxy 
	nanocomposite}
\label{section:Constitutive_model}

\begin{figure}[h!]
	\centering  
	\includegraphics[width=0.6\textwidth]{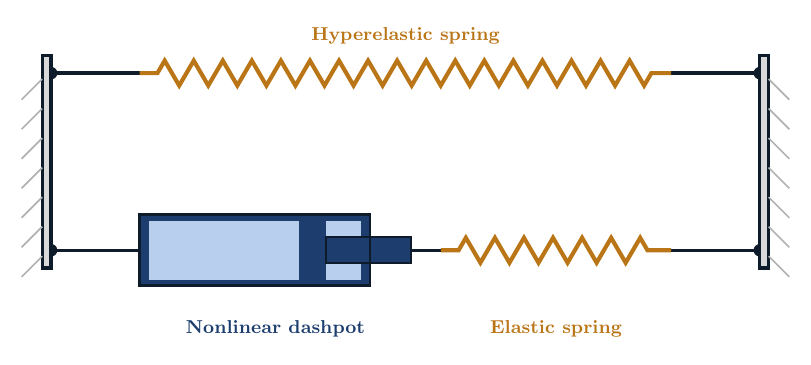}
	\caption{One-dimensional schematic representation of the 
		constitutive model.} 
	\label{rheological_model}
\end{figure}

The continuum formulation of a viscoelastic damage model 
\cite{arash2019viscoelastic} for nanoparticle/epoxy nanocomposites 
is briefly presented here. The rheological model 
(Fig.~\ref{rheological_model}) consists of a hyperelastic spring 
in parallel with a Maxwell element comprising a nonlinear dashpot 
and an elastic spring in series. The material response is expressed 
by (1) a hyperelastic spring in the case of large deformations, 
(2) a linearly elastic (Hencky model) spring considering the initial 
linear reaction, and (3) a nonlinear viscoplastic dashpot taking into 
account the strain-rate dependency of the material. A key feature of 
the present formulation is that the constitutive parameters governing 
both the elastic and viscoelastic responses are expressed as explicit 
functions of the BNP weight fraction $W$ and agglomerate size $S$, 
derived directly from large-scale CG simulations. This morphology 
dependence distinguishes the present model from prior formulations 
\cite{arash2019viscoelastic, unger2020effect} in which nanoparticle 
effects were either absent or captured only through a scalar 
amplification of the elastic response.

In this model, the total deformation gradient decomposes into elastic 
deformation $\textbf{F}_{e}$ (i.e. elastic spring) and inelastic 
$\textbf{F}_{i}$ (i.e. nonlinear dashpot).
\begin{equation}
	\begin{split}
		\textbf{F} = \textbf{F}_{e}\textbf{F}_{i}.
	\end{split}
	\label{Constitutive1}
\end{equation}


The overall free energy can be defined as an equilibrium 
(hyperelasticity) and a nonequilibrium (viscoelasticity) part,
\begin{equation}
	\begin{split}
		\psi(\textbf{C}, \textbf{F}_{i}, d) = 
		(1-d)[\psi_{eq}(\textbf{C}) + 
		\psi_{neq}(\textbf{F}_{i}^{-T}\textbf{C}\textbf{F}_{i}^{-1})],
	\end{split}
	\label{Constitutive2}
\end{equation}
where $\textbf{C}$ is the right Cauchy--Green tensor 
($\textbf{C} = \textbf{F}^{T}\textbf{F}$) and $d$ is a scalar 
damage variable in the range $[0,1)$ with initial value $d = 0$. 
In the part of the free energy regarding the equilibrium deformation, 
the eight-chain model \cite{arruda1993three} is used as,
\begin{equation}
	\begin{split}
		\psi_{eq} = nk\theta\left(\sqrt{N}\Lambda_{chain}\beta + 
		N \ln{\frac{\beta}{\sinh\beta}}\right),
	\end{split}
	\label{Constitutive3}
\end{equation}
\begin{equation}
	\begin{split}
		\beta = l^{-1}\left(\frac{\Lambda_{chain}}{\sqrt{N}}\right),
	\end{split}
	\label{Constitutive4}
\end{equation}
where $k$ is Boltzmann's constant, $n$ is the chain density, 
$\theta$ is temperature, $N$ is the number of rigid links between 
two crosslinks, and $l(x) = \text{coth}(x) - 1/x$ is the Langevin 
function.

The amplified chain stretch is defined as \cite{bergstrom1999mechanical},
\begin{equation}
	\begin{split}
		\Lambda_{chain} = \sqrt{X(W,S)\left(\lambda_{chain}^{2}-1\right)+1},
	\end{split}
	\label{Constitutive5}
\end{equation}
where $\lambda_{chain} = \sqrt{\bar{I}_{1}/3}$ is the respective 
stretch of each chain in the eight-chain model, with the first 
invariant $\bar{I}_{1} = \text{tr}[\bar{\textbf{B}}]$, the 
isochoric left Cauchy--Green tensor $\bar{\textbf{B}} = 
\bar{\textbf{F}}\bar{\textbf{F}}^{T}$, $\bar{\textbf{F}} = 
J^{-1/3}\textbf{F}$, and $J = \det[\textbf{F}]$. The quantity 
$X(W,S)$ is a morphology-dependent amplification factor that 
accounts for the effect of BNP weight fraction $W$ and agglomerate 
size $S$ on the nonlinear hyperelastic and linear elastic responses 
of the nanocomposite. A closed-form expression for $X(W,S)$ is 
derived from large-scale CG simulations in 
Sec.~\ref{subsection:Constitutive_nano}.

The nonequilibrium part $\psi_{neq}$ in Eq.~\ref{Constitutive2} 
is associated with the nonlinear viscoelastic deformation. For the 
elastic spring of the nonequilibrium part, the Hencky model is used 
\cite{henann2011large, xiao2002hencky},
\begin{equation}
	\begin{split}
		\psi_{neq} = X(W,S)\,\mu_{e}
		\left| \mathbf{E}'_e\right|^{2}    
		+ \frac{1}{2}X(W,S)\,\lambda_{e}
		\left(\text{tr}\,\textbf{E}_{e}\right)^{2},
	\end{split}
	\label{Constitutive6}
\end{equation}
where the logarithmic elastic strain 
$\textbf{E}_{e} = \ln(\textbf{V}_{e})$ and the deviatoric 
logarithmic elastic strain $\textbf{E}'_e$ are obtained   
from the polar decomposition 
$\textbf{F}_{e} = \textbf{V}_{e}\textbf{R}_{e}$, with 
$J_{e} = \det(\textbf{F}_{e})$ and $\textbf{R}_{e}$ the rotation 
tensor. The parameters $\mu_{e}$ and $\lambda_{e}$ are the 
Lam\'{e} moduli characterizing the linear material response.

The total Cauchy stress is defined by the damage-reduced sum of the 
equilibrium and nonequilibrium contributions as,
\begin{equation}
	\begin{split}
		\textbf{T} = (1-d)\left(\textbf{T}_{eq} + \textbf{T}_{neq}\right),
	\end{split}
	\label{Constitutive7}
\end{equation}
where $\textbf{T}_{eq}$ and $\textbf{T}_{neq}$ derive from the 
hyperelastic and viscoelastic free energy contributions, 
respectively \cite{sperling2005introduction},
\begin{equation}
	\begin{split}
		\textbf{T}_{eq} = \frac{X(W,S)\,\mu_{he}(\theta)}
		{J\Lambda_{chain}}
		\frac{l^{-1}\!\left(\dfrac{\Lambda_{chain}}{\sqrt{N}}\right)}
		{\dfrac{1}{\sqrt{N}}}\,\text{dev}[\bar{\textbf{B}}],
	\end{split}
	\label{Constitutive8}
\end{equation}
\begin{equation}
	\begin{split}
		\textbf{T}_{neq} = \frac{X(W,S)}{J_{e}}
		\left(2\mu_{e}(\theta)\,\textbf{E}_{e}^{0} + 
		\lambda_{e}(\theta)\,\text{tr}[\textbf{E}_{e}]\,\textbf{I}\right).
	\end{split}
	\label{Constitutive9}
\end{equation}
The explicit presence of $X(W,S)$ in both stress contributions 
ensures that the stiffening effect of BNPs on both the hyperelastic 
and viscoelastic responses is consistently captured through a single 
morphology-dependent quantity.

The time derivative of the inelastic deformation gradient is given by
\begin{equation}
	\begin{split}
		\dot{\textbf{F}}_{i} = \frac{\dot{\epsilon}_{i}}{\sigma_{neq}}
		\,\text{dev}[\textbf{T}_{neq}]\,\textbf{F}_{i},
	\end{split}
	\label{Constitutive10}
\end{equation}
where $\dot{\epsilon}_{i}$ is the viscoelastic flow rate and the 
Frobenius norm of the deviatoric driving stress is
\begin{equation}
	\begin{split}
		\sigma_{neq} = \left\|\text{dev}[\textbf{T}_{neq}]\right\|_{F}.
	\end{split}
	\label{Constitutive11}
\end{equation}
For the viscoelastic flow rate, the Argon model is used 
\cite{argon1977plastic},
\begin{equation}
	\begin{split}
		\dot{\epsilon}_{i} = \dot{\epsilon}_{0}(W,S)\exp\left[
		\frac{\Delta H(W,S)}{k\theta}
		\left(\left(\frac{\sigma_{neq}}{\sigma_{0}(W,S)}\right)^{5/6}
		-1\right)\right],
	\end{split}
	\label{Constitutive12}
\end{equation}
where $\dot{\epsilon}_{0}(W,S)$ is the pre-exponential factor, 
$\Delta H(W,S)$ is the activation energy, and $\sigma_{0}(W,S)$ 
is the athermal yield stress. All three Argon model parameters are 
expressed as explicit functions of the BNP weight fraction $W$ and 
agglomerate size $S$, derived from CG tensile simulations across a 
systematic range of nanoparticle morphologies. This morphology 
dependence of the viscoelastic flow parameters, together with the 
amplification factor $X(W,S)$, constitutes the principal extension 
of the present framework relative to prior models 
\cite{arash2019viscoelastic, unger2020effect}. The closed-form 
expressions for $\dot{\epsilon}_{0}(W,S)$, $\Delta H(W,S)$, and 
$\sigma_{0}(W,S)$ are presented.

Finally, the damage evolution law proposed by Qi 
\cite{qi2004constitutive} and Miehe \cite{miehe2000superimposed} 
is adopted as,
\begin{equation}
	\begin{split}
		\dot{d} = A(1-d)\Lambda_{chain}^{max},
	\end{split}
	\label{Constitutive13}
\end{equation}
where $A$ is the damage parameter and $\Lambda_{chain}^{max}$, 
as defined by Qi and Boyce \cite{qi2005stress}, is the maximum 
chain stretch attained over the deformation history,
\begin{equation}
	\begin{split}
		\Lambda_{chain}^{max} = \left\{\begin{matrix}
			0 & \Lambda_{chain} < \Lambda_{chain}^{max}\\[4pt]
			\Lambda_{chain} & \Lambda_{chain} \ge \Lambda_{chain}^{max}
		\end{matrix}\right..
	\end{split}
	\label{Constitutive14}
\end{equation}
The effect of temperature on the elastic and hyperelastic responses 
is accounted for through a modified Kitagawa model 
\cite{unger2020effect, kitagawa1977power}, which is applied to 
the reference Lam\'{e} moduli and the hyperelastic shear modulus as,
\begin{equation}
	\begin{split}
		\mu_{e}(\theta) &= \mu_{e,\text{ref}}
		\left(2-\exp\left[\alpha_{e}(\theta-\theta_{\text{ref}})\right]
		\right),\\[4pt]
		\lambda_{e}(\theta) &= \lambda_{e,\text{ref}}
		\left(2-\exp\left[\alpha_{e}(\theta-\theta_{\text{ref}})\right]
		\right),\\[4pt]
		\mu_{he}(\theta) &= \mu_{he,\text{ref}}
		\left(2-\exp\left[\alpha_{he}(\theta-\theta_{\text{ref}})\right]
		\right),
	\end{split}
	\label{Constitutive15}
\end{equation}
where $\mu_{e,\text{ref}}$, $\lambda_{e,\text{ref}}$, and 
$\mu_{he,\text{ref}}$ are the reference moduli of the pure epoxy 
at $\theta_{\text{ref}}$, and $\alpha_{e}$ and $\alpha_{he}$ are 
temperature scaling parameters for the elastic and hyperelastic 
responses, respectively. The nanoparticle contribution to the 
temperature-dependent stiffness enters through the amplification 
factor $X(W,S)$ in Eqs.~\ref{Constitutive8} and \ref{Constitutive9} 
rather than through additional morphology-dependent temperature 
parameters, which keeps the model compact while retaining physical 
consistency. It is worth noting that the Poisson's ratio is assumed 
to be temperature independent. The constitutive model is restricted 
to temperatures below the glass-transition region.

The constitutive equations are integrated numerically using an 
implicit fixed-point iteration scheme adapted from 
\cite{arash2026phase}. Given the deformation gradient 
$\mathbf{F}_{n+1}$ at the end of a time increment $\Delta t = 
t_{n+1} - t_n$, the algorithm iteratively updates the inelastic 
deformation gradient $\mathbf{F}^i$ until convergence. The 
nonlinear coupling between the inelastic deformation gradient and 
the non-equilibrium stress through the Argon flow rule in 
Eq.~\ref{Constitutive12} necessitates this iterative treatment: 
the flow rate $\dot{\epsilon}_i$ depends on $\sigma_{neq}$, which 
in turn depends on the elastic deformation 
$\mathbf{F}_e = \mathbf{F}_{n+1}(\mathbf{F}^i)^{-1}$ and hence 
on the current iterate of $\mathbf{F}^i$. Compared to the 
viscoelastic-viscoplastic algorithm in \cite{arash2026phase}, 
the present scheme is simplified by the absence of a viscoplastic 
mechanism: the iteration involves a single internal variable 
$\mathbf{F}^i$ rather than two coupled inelastic deformation 
gradients, and no yield threshold evaluation is required. The 
morphology-dependent parameters $X(W,S)$, $\dot{\epsilon}_0(W,S)$, 
$\Delta H(W,S)$, and $\sigma_0(W,S)$ depend only on the fixed 
material morphology and are therefore evaluated once before the 
iteration loop. The complete algorithm is summarized in 
Algorithm~\ref{alg:time_integration}.

\begin{algorithm}[H]
	\caption{Implicit time integration over $[t_n, t_{n+1}]$ 
		with fixed-point iteration \cite{arash2026phase}.}
	\label{alg:time_integration}
	\begin{algorithmic}[1]
		
		\State \textbf{Given:} $\mathbf{F}_{n+1}$, $\mathbf{F}^i_n$, 
		$d_n$, $\Lambda^{\max}_{chain,n}$, time step $\Delta t$, 
		morphology parameters $W$, $S$.
		
		\State \textbf{Precompute morphology-dependent parameters:}\vspace{-9pt}
		\[
		X = X(W,S), \quad
		\dot{\epsilon}_0 = \dot{\epsilon}_0(W,S), \quad
		\Delta H = \Delta H(W,S), \quad
		\sigma_0 = \sigma_0(W,S).
		\]
		
		\State \textbf{Initialize:} $\mathbf{F}^i = \mathbf{F}^i_n$, 
		tolerance $\varepsilon_{\text{tol}}$, 
		\texttt{converged} $= \text{false}$.
		
		\While{\texttt{converged} $= \text{false}$}
		
		\State \textbf{Kinematics:}\vspace{-9pt}
		\[
		J = \det(\mathbf{F}_{n+1}), \quad
		\mathbf{F}_e = \mathbf{F}_{n+1}\,(\mathbf{F}^i)^{-1}, \quad
		\mathbf{F}_e = \mathbf{V}_e \mathbf{R}_e, \quad
		\mathbf{E}_e = \ln(\mathbf{V}_e), \quad
		J_e = \det(\mathbf{F}_e).
		\]
		\vspace{-24pt}
		\State \textbf{Compute non-equilibrium stress
			(Eq.~\eqref{Constitutive9}):}\vspace{-9pt}
		\[
		\mathbf{T}_{neq} = \frac{X(W,S)}{J_e}
		\left[2\mu_e(\theta)\,\mathbf{E}'_e 
		+ \lambda_e(\theta)\,\mathrm{tr}[\mathbf{E}_e]\,\mathbf{I}\right].
		\]
		\vspace{-24pt}
		\State \textbf{Compute viscoelastic flow rate 
			(Eq.~\eqref{Constitutive12}):}\vspace{-9pt}
		\[
		\dot{\epsilon}_i = \dot{\epsilon}_0(W,S)\exp\!\left[
		\frac{\Delta H(W,S)}{k\theta}
		\left(\left(\frac{\sigma_{neq}}{\sigma_0(W,S)}\right)^{5/6}
		- 1\right)\right].
		\]
		\vspace{-24pt}		
		\State \textbf{Compute inelastic deformation gradient update 
			(Eq.~\eqref{Constitutive10}):}\vspace{-9pt}
		\[
		\mathbf{D}^i = \frac{\dot{\epsilon}_i}{\sigma_{neq}}
		\,\mathrm{dev}[\mathbf{T}_{neq}], \qquad
		\mathbf{F}^i_{\text{new}} = 
		\exp\!\left(\Delta t\,\mathbf{D}^i\right)\mathbf{F}^i_n.
		\]
		
		\State \textbf{Convergence check:}
		$
		\Delta^i = \|\mathbf{F}^i_{\text{new}} - \mathbf{F}^i\|_F.
		$
		\If{$\Delta^i < \varepsilon_{\text{tol}}$}
		\State $\mathbf{F}^i_{n+1} \leftarrow 
		\mathbf{F}^i_{\text{new}}$, \quad
		\texttt{converged} $= \text{true}$.
		\Else
		\State $\mathbf{F}^i \leftarrow \mathbf{F}^i_{\text{new}}$.
		\EndIf
		
		\EndWhile
		
		\State \textbf{Compute equilibrium stress (Eq.~\eqref{Constitutive8}) using}\vspace{-9pt}
		\[
		\bar{\mathbf{F}} = J^{-1/3}\mathbf{F}_{n+1}, \quad
		\bar{\mathbf{B}} = \bar{\mathbf{F}}\bar{\mathbf{F}}^T, \quad
		\lambda_{chain} = \sqrt{\bar{I}_1/3},
		\]
		\vspace{-24pt}		
		\State \textbf{Damage update (Eqs.~\ref{Constitutive13} 
			and \ref{Constitutive14}):}\vspace{-9pt}
		\[
		\Lambda^{\max}_{chain,n+1} = 
		\max\!\left(\Lambda^{\max}_{chain,n},\,\Lambda_{chain}\right),
		\]
		\vspace{-24pt}
		\[
		d_{n+1} = 1 - (1 - d_n)
		\exp\!\left(-A\,\Lambda^{\max}_{chain,n+1}\,\Delta t\right).
		\]
		\vspace{-24pt}		
		\State \textbf{Compute total Cauchy stress (Eq.~\eqref{Constitutive7}):}\vspace{-9pt}
		\[
		\mathbf{T}_{n+1} = 
		(1 - d_{n+1})\left(\mathbf{T}_{eq} + \mathbf{T}_{neq}\right).
		\]
		
	\end{algorithmic}
\end{algorithm}

\begin{figure}[h!]
\centering
\begin{tikzpicture}[
	node distance = 0.55cm and 1.2cm,
	every node/.style = {font=\scriptsize},
	block/.style = {
		rectangle, rounded corners=3pt,
		minimum width=3.8cm, minimum height=0.75cm,
		text centered, text width=3.6cm,
		draw=#1!80!black, fill=#1!18,
		line width=0.6pt
	},
	blockW/.style = {
		rectangle, rounded corners=3pt,
		minimum width=8.4cm, minimum height=0.75cm,
		text centered, text width=8.2cm,
		draw=#1!80!black, fill=#1!18,
		line width=0.6pt
	},
	titleblock/.style = {
		rectangle, rounded corners=3pt,
		minimum width=3.8cm, minimum height=0.7cm,
		text centered, text width=3.6cm,
		draw=#1!80!black, fill=#1!60,
		text=white, font=\small\bfseries,
		line width=0.6pt
	},
	arrow/.style = {
		->,>=Stealth,
		line width=0.8pt,
		color=#1
	},
	groupbox/.style = {
		rectangle, rounded corners=5pt,
		draw=#1!70!black, fill=#1!6,
		line width=0.8pt, dashed
	}
	]
	
	\definecolor{cgcol}{RGB}{0,53,107}      
	\definecolor{expcol}{RGB}{179,128,0}    
	\definecolor{modelcol}{RGB}{60,60,60}   
	\definecolor{valcol}{RGB}{0,110,60}     
	
	\node[titleblock=cgcol] (cg0)
	{\scriptsize CG Simulation Stream};
	
	\node[block=cgcol, below=of cg0] (cg1)
	{Molecular model construction \& curing simulation};
	
	\node[block=cgcol, below=of cg1] (cg2)
	{CG tensile simulations\\
		(varying $\dot{\varepsilon}$, $T$, $W$, $S$)};
	
	\node[block=cgcol, below=of cg2] (cg3)
	{Yield stress extraction\\
		via cubic spline fit};
	
	\node[block=cgcol, below=of cg3] (cg4)
	{Argon model parameters\\
		$\dot{\epsilon}_0(W,S)$,\;
		$\Delta H(W,S)$,\;
		$\sigma_0(W,S)$};
	
	\node[block=cgcol, below=of cg4] (cg5)
	{Amplification factor\\
		$X(W,S)$};
	
	\node[titleblock=expcol, right=of cg0] (ex0)
	{\scriptsize Experimental Stream};
	
	\node[block=expcol, below=of ex0] (ex1)
	{BNP/epoxy specimen\\
		fabrication};
	
	\node[block=expcol, below=of ex1] (ex2)
	{Tensile tests\\
		(two $\dot{\varepsilon}$, three $T$,\\
		pure epoxy)};
	
	\node[block=expcol, below=of ex2] (ex3)
	{Genetic algorithm\\
		parameter calibration};
	
	\node[block=expcol, below=of ex3] (ex4)
	{Hyperelastic parameters\\
		$\mu_{he,\mathrm{ref}}$,\;
		$\alpha_{he}$,\; $N$};
	
	\node[block=expcol, below=of ex4] (ex5)
	{Elastic \& damage parameters\\
		$\mu_{e,\mathrm{ref}}$,\;
		$\lambda_{e,\mathrm{ref}}$,\;
		$\alpha_{e}$,\; $A$};
	
	\node[blockW=modelcol,
	below=1.1cm of $(cg5.south)!0.5!(ex5.south)$]
	(model)
	{Fully parametrized constitutive model\\[2pt]
		$\mathbf{T} = (1-d)(\mathbf{T}_{eq} + \mathbf{T}_{neq})$};
	
	\node[blockW=valcol, below=0.55cm of model] (val)
	{Independent experimental validation\\
		(BNP 5, 10, 15\,wt\%, varying $\dot{\varepsilon}$ and $T$,
		excluded from calibration)};
	
	\begin{scope}[on background layer]
		\node[groupbox=cgcol,
		fit=(cg0)(cg1)(cg2)(cg3)(cg4)(cg5),
		inner sep=6pt] (cgbox) {};
		\node[groupbox=expcol,
		fit=(ex0)(ex1)(ex2)(ex3)(ex4)(ex5),
		inner sep=6pt] (expbox) {};
	\end{scope}
	
	\foreach \from/\to in {cg0/cg1, cg1/cg2, cg2/cg3,
		cg3/cg4, cg4/cg5}
	\draw[arrow=cgcol] (\from) -- (\to);
	
	\foreach \from/\to in {ex0/ex1, ex1/ex2, ex2/ex3,
		ex3/ex4, ex4/ex5}
	\draw[arrow=expcol] (\from) -- (\to);
	
	\draw[arrow=cgcol]   (cg5.south)  -- ++(0,-0.45)
	-| (model.north west);
	\draw[arrow=expcol]  (ex5.south)  -- ++(0,-0.45)
	-| (model.north east);
	
	\draw[arrow=modelcol] (model) -- (val);
	
\end{tikzpicture}
\caption{Flowchart of the proposed simulation-informed constitutive
	modeling framework. The CG simulation stream (left, blue) provides
	the morphology-dependent amplification factor $X(W,S)$ and the
	Argon viscoelastic model parameters $\dot{\epsilon}_0(W,S)$,
	$\Delta H(W,S)$, and $\sigma_0(W,S)$ as functions of
	BNP weight fraction $W$ and agglomerate size $S$. The experimental
	stream (right, gold) supplies the hyperelastic, elastic, and damage
	parameters of the pure epoxy through a genetic algorithm calibration. The two streams converge into a fully parametrized constitutive model, whose predictive capability is assessed against independent	experimental data not used in the calibration process.}
\label{fig:flowchart}
\end{figure}
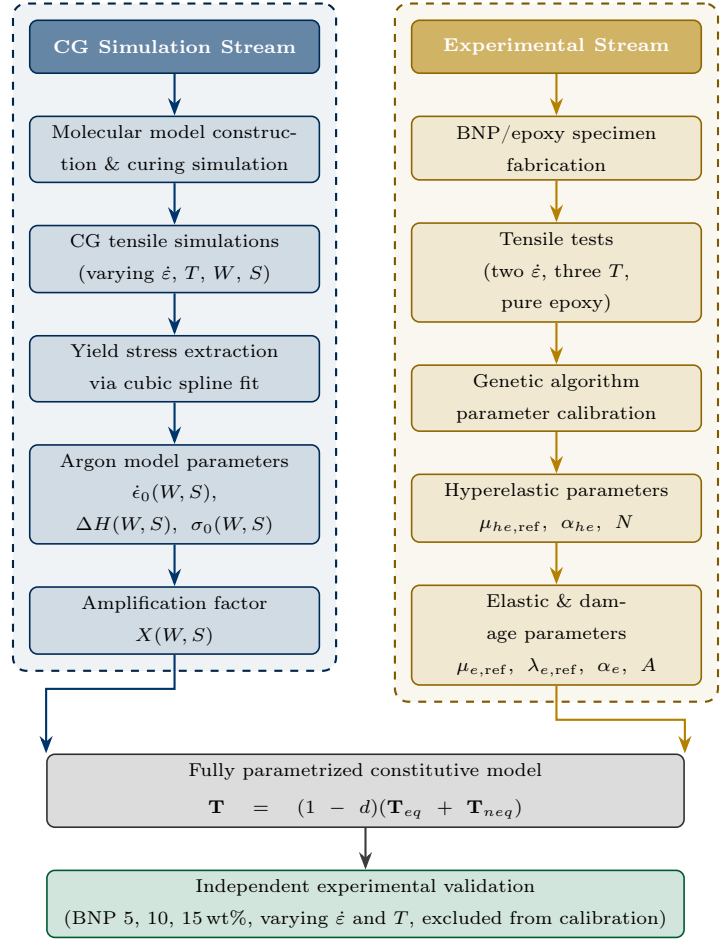

Fig.~\ref{fig:flowchart} summarizes the overall parameter 
identification strategy of the proposed framework. The framework 
combines two complementary streams. In the CG simulation stream, 
large-scale tensile simulations of BNP/epoxy systems at systematically 
varied strain rates, temperatures, weight fractions $W$, and 
agglomerate sizes $S$ are used to extract the morphology-dependent 
amplification factor $X(W,S)$ and the Argon viscoelastic model 
parameters $\dot{\epsilon}_0(W,S)$, $\Delta H(W,S)$, and 
$\sigma_0(W,S)$ as analytical functions. In the 
experimental stream, tensile tests on pure epoxy specimens at two 
strain rates and three temperatures are used to calibrate the 
remaining constitutive parameters, namely the hyperelastic shear 
modulus $\mu_{he,\mathrm{ref}}$ and network parameters $\alpha_{he}$ 
and $N$, the elastic Lam\'{e} moduli $\mu_{e,\mathrm{ref}}$ and 
$\lambda_{e,\mathrm{ref}}$ with their temperature scaling $\alpha_e$, 
and the damage parameter $A$, through a genetic algorithm 
optimization. The two streams together fully parametrize the 
constitutive model, which is subsequently validated against 
independent experimental data for BNP/epoxy nanocomposites at 
weight fractions and conditions not included in the calibration 
process. This structure significantly reduces the experimental 
effort required for model development while maintaining 
predictive accuracy across a broad morphological and 
thermomechanical parameter space.


\section{Material system}
\label{section:Material}

In the present study, the PNC is a low-viscosity infusion system utilized for the infusion process in the big and thick composite parts, such as wind turbine blades. The molecular structure of the epoxy resin and boehmite nanoparticles are explained in the following. 

\subsection{Epoxy resin}

\begin{figure}[h!]
	\centering
	\begin{subfigure}[b]{0.60\linewidth}
		\centering
		\includegraphics[width=\textwidth]{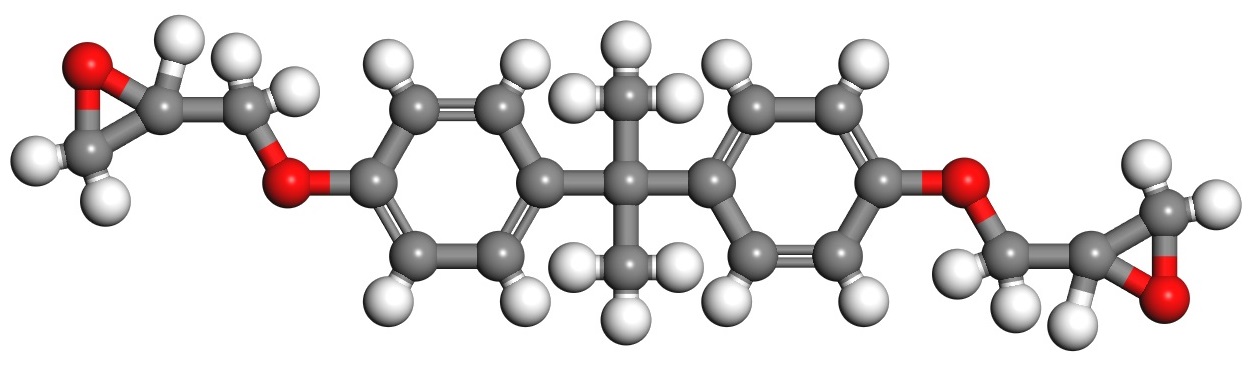}
		\caption{}
		\label{Material_system0}
	\end{subfigure}
	\begin{subfigure}[b]{0.35\linewidth}
		\centering
		\includegraphics[width=\textwidth]{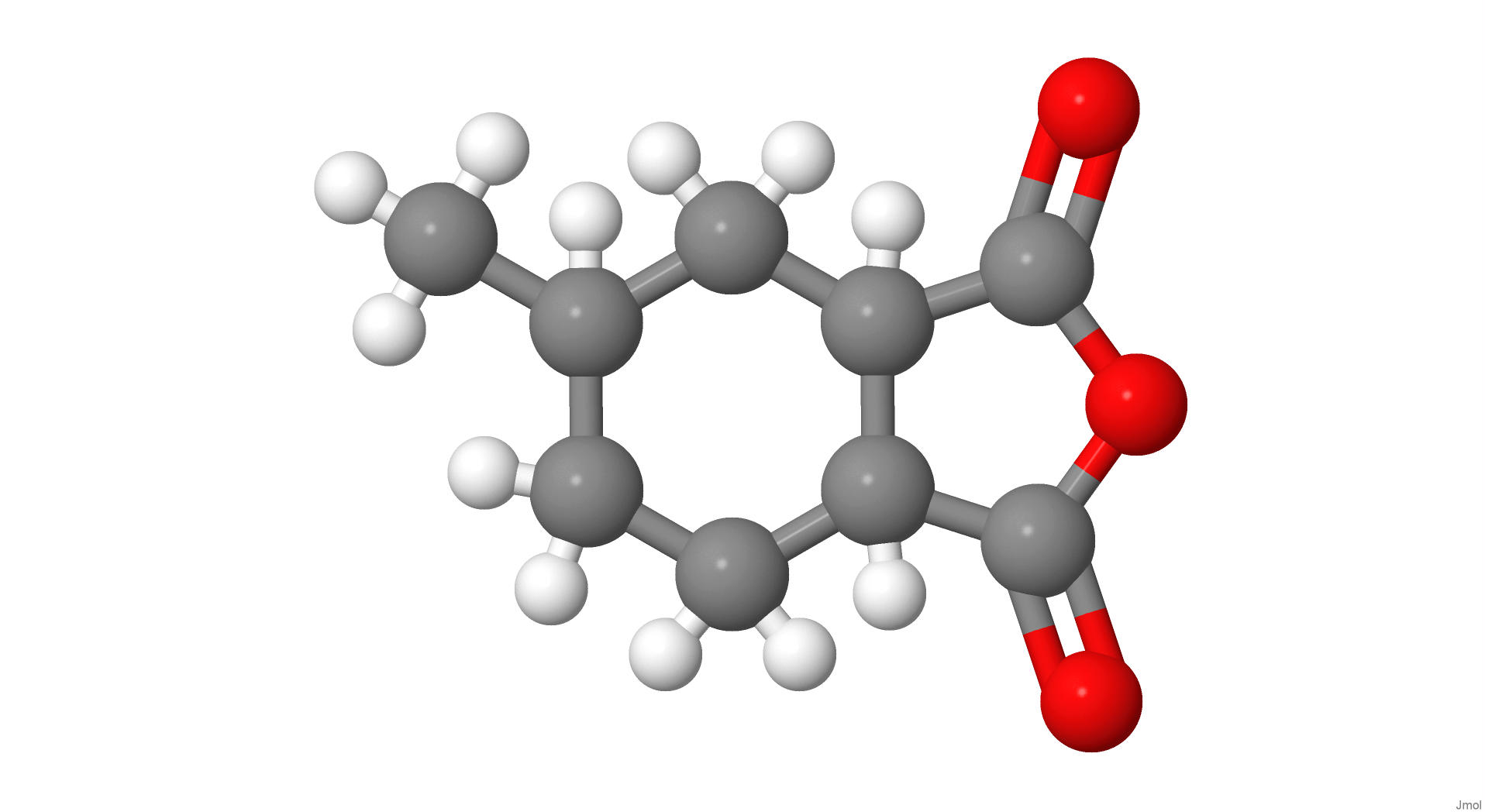}
		\caption{}
		\label{Material_system1}  
	\end{subfigure}  
	\caption{Molecular structures of (a)  a bisphenol-A-diglycidylether monomer and (b) a 4-methyl-1,2-cyclohexanedicarboxylic anhydride curing agent. The red, grey and white colors represent oxygen, carbon and hydrogen atoms, respectively.} 
	\label{Material_system}
\end{figure}

\begin{figure}[h!]
 	\centering
 	\begin{subfigure}[b]{0.65\linewidth}
 		\centering
 		\includegraphics[width=\textwidth]{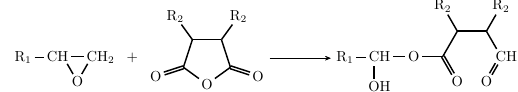}
 	\end{subfigure}   
 	\begin{subfigure}[b]{0.75\linewidth}
 		\centering
 		\includegraphics[width=\textwidth]{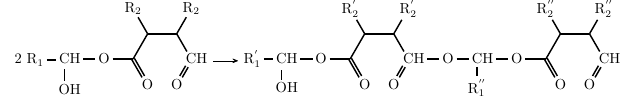}
 	\end{subfigure}   
 	\caption{Curing reaction mechanism between DGEBA epoxy and anhydride curing agent.} 
 	\label{curing_mechanism}
\end{figure}

The matrix consists of bisphenol-A-diglycidylether (DGEBA), and is cured with 4-methyl-1,2-cyclohexane dicarboxylic anhydride (MTHPA). The molecular structure of DGEBA and the curing agent are shown in Figs.~\ref{Material_system0} and \ref{Material_system1}. The reaction opens the epoxide ring and hydrolyzes the curing agent, leading to the formation of carboxylic acid groups. In the next step, the carboxylic acids react with epoxy groups and produce hydroxyl groups to react more with DGEBA and MTHPA, as shown in Fig.~\ref{curing_mechanism}. The mixing ratio is the standard stoichiometric epoxy monomer/hardener mixing ratio which is equal to 100:90~\cite{jux2018mechanical, khorasani2019effect}.

\begin{figure}[h!]
	\centering
	\begin{tikzpicture} 
	\node[inner sep=0pt]  at (0.0,0.0) {\includegraphics[width=0.7\textwidth]{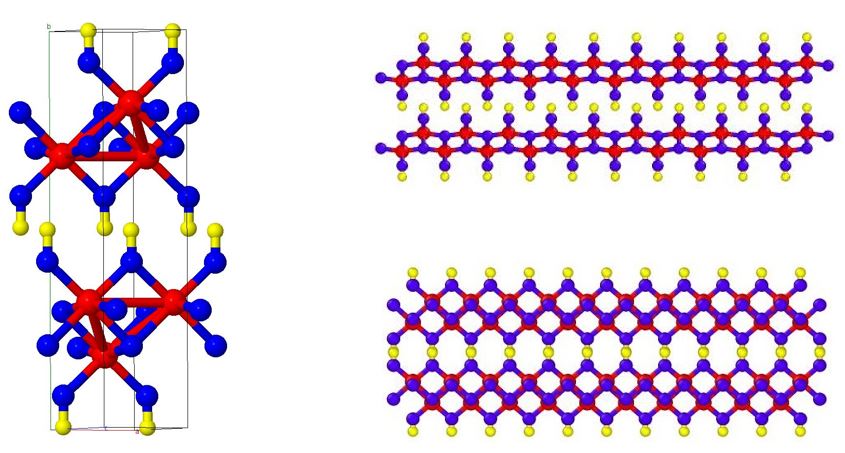}};
	\node (A) at (-4.4, 2.5) {$b$};
	\node (A) at (-4.0, -2.3) {$c$};
	\node (A) at (-3.65, -2.4)  {$a$};
	\node (A) at (-3.5, -2.85)  {$(a)$};
	\node (A) at (2, -2.85)  {$(c)$};
	\node (A) at (2, 0.2)  {$(b)$};
	\end{tikzpicture}
	\caption{Molecular structures of (a) the unit cell of the crystalline structure of boehmite. (b) X-Y direction and (c) Y-Z direction of two layer boehmite structure. The color red, blue and yellow represent aluminum, oxygen and hydrogen atoms, respectively.} 
	\label{Material_system_bohi}
\end{figure}

\subsection{Nanoparticles}

Nanocomposite systems contain nano-scaled reinforcement material to enhance mechanical, thermal, and electrical properties. In the present study, boehmite nanoparticles (BNP) are added to our epoxy system. Boehmite is an aluminum oxide hydroxide ($\gamma$-AlO(OH)) with lattice parameters a = 3.693, b = 12.221, and c = 2.865~$\text{\AA}$. As illustrated in Fig.~\ref{Material_system_bohi}, central aluminum atoms are bonded to double layers of oxygen to produce the crystalline structure of BNPs. The oxygen in the out-face is bonded by hydrogen bonds to the hydroxyl group of the neighboring layer of octahedrons.


\section{Coarse-grained modeling}
\label{section:cg}

\subsection{Mapping scheme}
\label{subsection:Mapping}

In coarse-graining, a set of atoms is mapped into a CG super-atom called a bead. The first step in coarse-graining is to define a mapping scheme that makes the model capable of keeping the identity of the material chemistry. Mapping schemes can be decided based on the chemical compositions of molecules and repeating units of the monomers. In this work, the highest level of coarse-graining is selected due to the discussions explained in \cite{hente2021optimization}.

\begin{figure}[h!]
	\centering
	\begin{subfigure}[b]{0.5\linewidth}
		\centering
		\begin{tikzpicture}
		\node[inner sep=0pt]  at (0.0,0.0) 	
		{\includegraphics[width=6.5cm,clip=false]{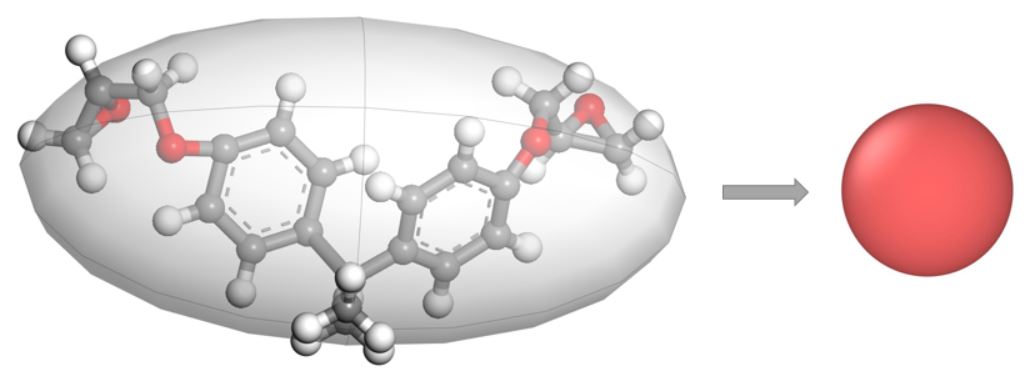}};
		\node (A) at  (2.6, 0.75)  {$A$};
		\end{tikzpicture}
		\caption{}
		\label{CGMD_A}  
	\end{subfigure}   
	\begin{subfigure}[b]{0.4\linewidth}
		\centering
		\begin{tikzpicture}
		\node[inner sep=0pt]  at (0.0,0.0) 	
		{\includegraphics[width=5.0cm,clip=false]{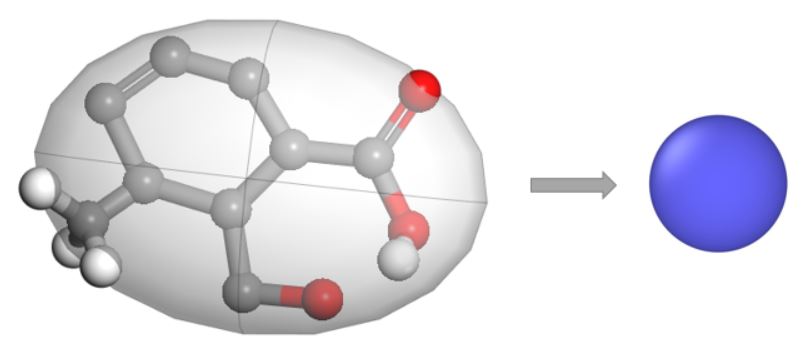}};
		\node (A) at  (2, 0.6)  {$B$};
		\end{tikzpicture}
		\caption{}
		\label{CGMD_B}  
	\end{subfigure}   
	\caption{(a) A DGEBA epoxy monomer for all-atom representation and its corresponding A CG bead, (b) An anhydride hardener monomer for all-atom representation and its corresponding B CG bead.} 
	\label{CGMD_AB}
\end{figure}   

\subsubsection{Epoxy resin}

To  introduce a high level of coarse-graining, each monomer is mapped into one bead as illustrated in Figs.~\ref{CGMD_A} and \ref{CGMD_B}. A CG system consists of two types of beads, one for the bisphenol-A monomer denoted by A and one for the anhydride hardener denoted by B. The atomic mass of A and B beads are 340.4128 and 166.1739~amu, respectively. In this case, the DOF decreases by 49 and 22 times for beads A and B, respectively. 

\subsubsection{Nanoparticles}

\begin{figure}[h!]
	\centering
	\begin{subfigure}[b]{0.8\linewidth}
		\centering
		\begin{tikzpicture}
		\node[inner sep=0pt]  at (0.0,0.0) 	
		{\includegraphics[width=8.0cm,clip=false]{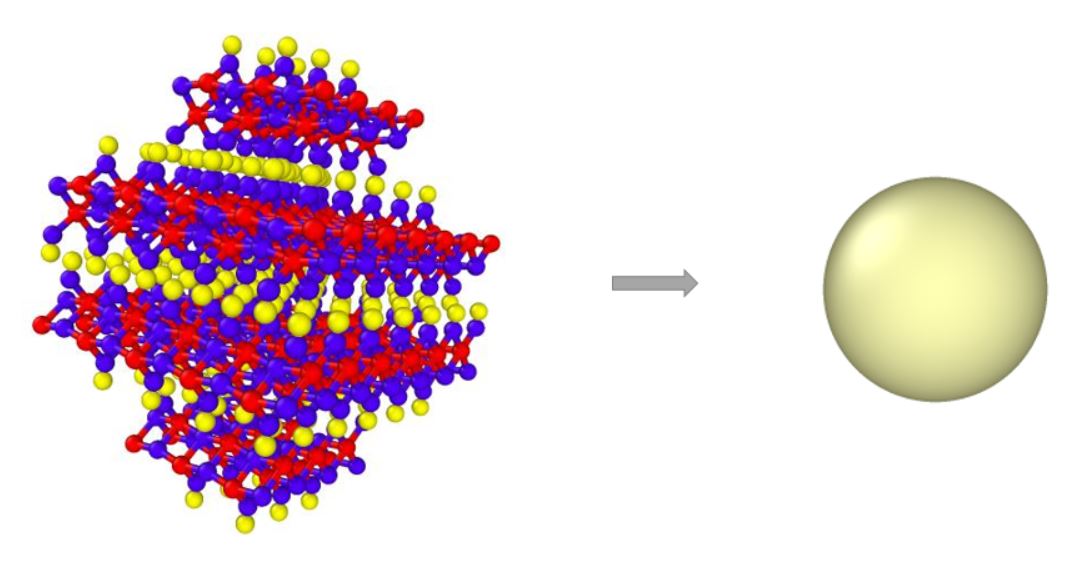}};
		\node (A) at  (3, 1)  {$P$};
		\end{tikzpicture}
	\end{subfigure}   
	\caption{Four layer boehmite structure with the size of 20~$\text{\AA}$ and its corresponding CG bead denoted by P.} 
	\label{BohmiteCG}
\end{figure}

The bulk modulus of boehmite is 93~GPa \cite{tunega2011theoretical}. The structural properties have been studied experimentally through XRD \cite{bokhimi2001relationship} and Raman spectroscopy \cite{kiss1980raman} and numerically through quantum mechanics \cite{tunega2011theoretical,noel2009ab}. 
The high mechanical modulus of BNPs compared to the epoxy matrix allows us to model them as rigid particles in CG modeling. In the following simulations, one primary particle with 20~$\text{\AA}$ length is mapped to one CG bead named P. In this study, commercially spray-dried boehmite nanoparticles with the shape orthorhombic as primary particles are used (DISPERAL HP14, SASOL). Compared to its full atomistic system, the number of DOF decreases by 656 times in this CG model. The atomic mass of P is 10433~amu. The primary particle is illustrated in Fig.~\ref{BohmiteCG}.  

\subsection{Coarse-grained potentials}

The total potential energy of the system is the summation of energy terms associated with bond and nonbond potential functions \cite{arash2015mechanical} given by
\begin{equation}
E_{total}(d,\theta,r)= \sum_{i}E_{b_{i}}+ \sum_{j}E_{a_{j}}+ \sum_{k}E_{nonb_{k}},
\label{Total_eng}
\end{equation}
where $E_{b}$, $E_{a}$ are the terms of energy corresponding to the variation of the bond length and angle, respectively, while $E_{nonb}$ stands for the nonbond interactions.  

The potential used in this CG model for stretching is the Morse potential~\cite{mayo1990generic}:
\begin{equation}
\begin{split}
E_{b}(r)=D_{b}(1-e^{-\alpha(r-r_{0})})^{2},
\end{split}
\label{Bond_Morse}
\end{equation}
where $r_{0}$ is the equilibrium bond distance, $\alpha$ is a stiffness parameter, and $D_{b}$ shows the depth of the potential well. Considering the bond between epoxy and BNPs~\cite{fankhanel2019elastic}, there are two types of bonds in our nanocomposite system, AB and PB. 
 
The harmonic force field is applied for the bending potential in this CG model:
\begin{equation}
\begin{split}
E_{a}(\theta)= \frac{K_{\theta}}{2}(\theta-\theta_{0})^{2},
\end{split}
\label{Angle_harmonic}
\end{equation}
where $K_{\theta}$ and $\theta_{0}$ represent the spring constant and the equilibrium angle, respectively, two types of angles in the CG model are BAB and PBP.  

The nonbond interactions are modeled using a Lennard-Jones-(12,6) potential with a cut-off distance of 20~$\text{\AA}$,
\begin{equation}
E_{vdW}(r)=\epsilon\left[(\frac{ \sigma }{r})^{12}- (\frac{ \sigma }{r})^{6}\right],
\label{LJ}
\end{equation}
where $\epsilon$ is the equilibrium well depth and $\sigma$ is the equilibrium distance. Since there are three CG particle types, the CG model of the nanocomposite contains six unknown non-bonded force-field parameters: $\epsilon^{AA}$, $\epsilon^{BB}$, $\epsilon^{PP}$, $\sigma^{AA}$, $\sigma^{BB}$, $\sigma^{PP}$. 

For nonbond interactions between unlike bead pairs, the arithmetic mixing rule~\cite{lebowitz1971mixtures} is used to obtain pair potentials, where the Lennard-Jones parameters are given by $\epsilon_{ij}~=~\sqrt{\epsilon_{i}\epsilon_{j}}$ and $\sigma_{ij}~=~\frac{1}{2}(\sigma_{i} + \sigma_{j})$. The force field parameters taken from Hente et al.~\cite{hente2025enhancement} are listed in Table~\ref{force-field}. The parameters have been calibrated using an artificial neural network-assisted optimization method. The method has been suggested as an efficient and robust algorithm for calibrating CG force fields with high accuracy and low computational cost, allowing thermo-mechanical properties estimation for BNP/epoxy nanocomposites over a broad range of temperatures.

\begin{table}[h]
	\caption{Force field parameters of the CG model.}
	\label{force-field}
	\centering
	\begin{tabular}{lcccc}
		\toprule
		Type of interactions & Parameters & \multicolumn{2}{c}{Epoxy} & BNP \\
		\midrule
		Bond & $D_{b}$ (kcal/mol/\AA$^{2}$) & \multicolumn{2}{c}{$45.86$} & $95.81$ \\
		& $\alpha$ (-) & \multicolumn{2}{c}{$0.97$} & $0.98$ \\
		& $r_{0}$ (\AA) & \multicolumn{2}{c}{$5.04$} & $12.05$ \\
		Angle & $K_{\theta}$ (kcal/mol/rad$^{2}$) & \multicolumn{2}{c}{$25.07$} & $98.23$ \\
		& $\theta_{0}$ ($^{\circ}$) & \multicolumn{2}{c}{$149.74$} & $150.98$ \\
		\cmidrule(lr){3-4}
		& & Bead A & Bead B & \\
		\cmidrule(lr){3-4}
		vdW & $\epsilon$ (kcal/mol) & $4.53$ & $4.47$ & $100.64$ \\
		& $\sigma$ (\AA) & $7.94$ & $7.78$ & $16.19$ \\
		\bottomrule
	\end{tabular}
\end{table}

%
%
%
%
%
%
%
%

\section{Experiments}
\label{section:ExP_explain}

\subsection{Manufacturing}

\begin{figure}[h!]
	\centering
	\includegraphics[width=0.7\textwidth]{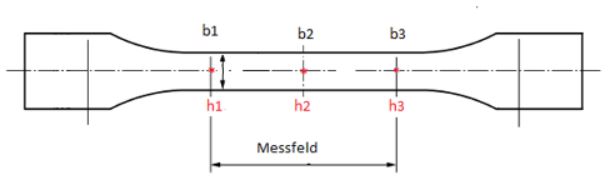}
	\caption{Schematic picture of dog-bone test specimen prepared for the testing.} 
	\label{tensile_exp1}
\end{figure}

Manufacturing the BNP/epoxy nanocomposite test specimens starts by mixing the nanoparticles with the epoxy system. The BNP nanoparticles are processed in suspensions with high viscosity and mass fractions of 30 wt\%. A kneader was used for dispersing nanoparticles which applied shear stressing between surfaces and laminar flow. After BNP dispersion, the degree of agglomeration was verified using a scanning electron microscope (SEM) and dynamic light scattering (DLS) \cite{jux2017effects}. According to the guidelines of the manufacturers, the final nanocomposite system is composed of an epoxy-hardener mixing ratio of 100:90~\cite{jux2018mechanical, khorasani2019effect} using dispersed nanoparticles. Highly viscous suspensions are blended with DGEBA using a vacuum centrifugal mixer. This process is then repeated at least two times with rotational speeds up to 2100 rpm to prepare a high-quality mixture. To reduce the trapped air inside the mixture, it is subjected to vacuum up to 3 mbar. The hardener and accelerator are then added to the mixture. Performing visual tests after each blending step aids in confirming the quality of the mixture. For the manufacturing of test specimens, the mixture is subjected to a pre‐heated tool (80 $^{\circ}$C) while getting cured by a water-based mold release system (WaterWorks Aerospace Release). The mixture undergoes gelation at 80 $^{\circ}$C for four hours, followed by post-curing at 120 $^{\circ}$C for four hours. A milling process is then utilized in the final step to prepare the test specimens by cutting the cured plates into a standardized shape according to DIN EN ISO 527, which is a dog-bone shape as illustrated in Fig.~\ref{tensile_exp1}.

\begin{figure}[h!]
	\centering
	\includegraphics[width=0.7\textwidth]{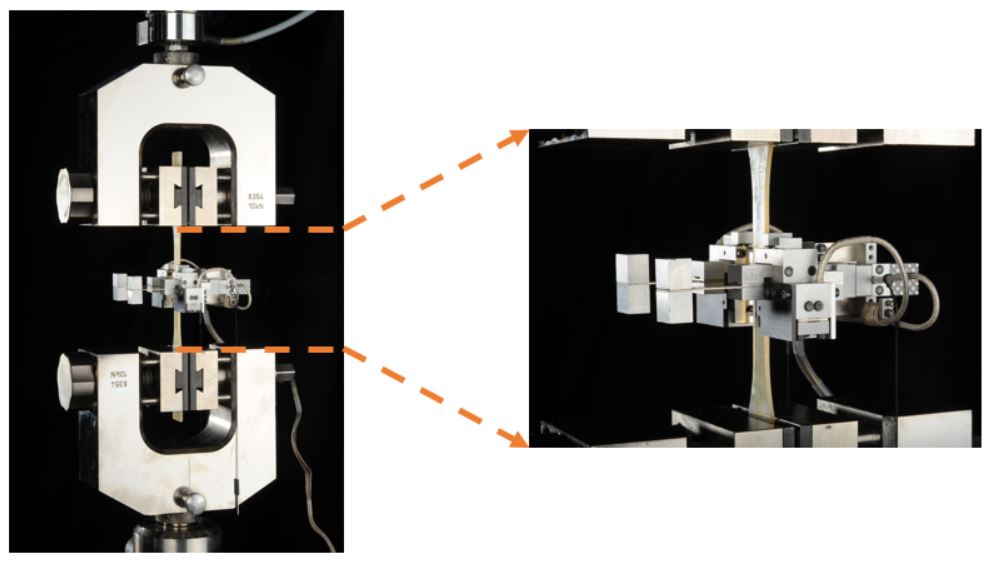}
	\caption{Tensile test device with extensometer until failure.} 
	\label{tensile_exp2}
\end{figure}

\subsection{Mechanical tests}

The experimental test results are statistically applicable with ten different specimens for the tensile tests. The specimens were conditioned for at least two days at room temperature and humidity of 51\%. For mechanical testing, a Zwick/Roell (Z005) static testing machine is utilized, as shown in Fig.~\ref{tensile_exp2}. Specimens with a thickness of 2 mm are produced for the tensile tests. They are tested according to testing standard DIN EN ISO 527‐1 and DIN EN ISO 527‐2, sample type 1B. The tensile tests are conducted with two test speeds of 0.1 mm/min and 10 mm/min while the initial load is kept constant and equal to 5 N. The stress–strain curves are then recorded with these conditions. The effect of nanoparticle content on the tensile properties was studied using the standardized test configuration for weight fractions of 5\%, 10\%, and 15\%. To study the effect of temperature, two more configurations were tested at higher temperatures of 40 and 80~$^{\circ}$C, with the test speed of 1 mm/min and 50\% relative humidity. 
For the configurations tested at elevated temperatures of 40 and 80~$^\circ$C, the specimens were placed in a climate chamber and kept at the target temperature for at least five minutes prior to testing to ensure a uniform temperature distribution throughout the specimen.

\section{Results and discussion}
\label{section:Results}

The following section presents CG simulations of the epoxy system to construct realistic molecular models of cured epoxies. Next, a CG simulation-based method to identify the material parameters associated with the nonlinear viscoelastic dashpot element is presented. Further to the simulation measured parameters, those of the elastic and hyperelastic springs and the damage variable are obtained through a calibration process with experimental data for the pure epoxy system. Also, an amplification factor taking into account the effect of nanoparticles on the material behavior of BNP/epoxy nanocomposites is calibrated using CG simulations. Finally, the  capability of the calibrated constitutive model to predict the stress–strain response of BNP/epoxy nanocomposites is evaluated using experimental data.

\subsection{Coarse-grained simulations} 
\label{subsection:CGSim} 


The non-cross-linked models are generated by placing DGEBA and hardener molecules randomly in a periodic simulation box with an initial density of 1.2 gr/cc. As mentioned earlier, the mixing ratio of epoxy monomer/hardener is 100:90. The non-cross-linked models are produced using an open-source package, PACKMOL~\cite{martinez2009software}. In the following simulations, the Nos\'{e}-Hoover thermostat and barostat~\cite{melchionna1993hoover} is used for the system temperature and pressure conversions. 

In the curing process, epoxy monomers connect to an agent molecule, as explained in Section~\ref{section:Material}. In atomistic, the connection for the epoxy system occurs by the formation of the methyl group of the monomer, which links to the hydroxyl group of the agent molecule. In the BNP/epoxy nanocomposite, the hydroxyl groups of the boehmite surface can also participate in the curing reaction~\cite{fankhanel2019elastic}. To prepare the same cross-linked nanocomposite system, we introduce two bond types, epoxy-agent (AB) and BNP-agent (PB).

For the cross-linking simulation of an uncured system, after energy minimization, the temperature is linearly increased to the curing temperature, 450 K, ~\cite{jux2018mechanical}. 
The system is equilibrated under NPT ensemble for 1~ns at a constant pressure of 1~atm. The following steps are then conducted for the cross-linking simulation. First, a bond between with the shortest possible length, with a cut-off of 10~$\text{\AA}$ for the reaction distance, is formed for the epoxy (the distance can not be less than 4~$\text{\AA}$). Next, an equilibration is performed for a time period of 2.5 ps
 
These two steps are repeated until the final degree of curing of around 90~\% is reached. The cut-off for the first step in the case of BNP/epoxy nanocomposite is 16~$\text{\AA}$ with the same minimum distance  of 4~$\text{\AA}$.
The time step was 2 fs during the simulation. The duration of each step and the cut-off value was selected in such a way as to ensure a well-equilibrated system and a reasonable computing time. 
The curing process is conducted by the Large-scale Atomic/Molecular Massively Parallel Simulator (LAMMPS)~\cite{plimpton1995fast} with our modification to fix the bond/create command. The proposed modification works in such a way that the distances between all possible reaction groups are calculated, and the bond with the smallest reaction distance is formed. This means that all reaction distances larger than the chosen cut-off is discarded. For visualization and analysis of the data, OVITO~\cite{stukowski2009visualization} is used in this work.


After cross-linking, the cured system is cooled down to room temperature and then relaxed using NPT for 1.5 ns. The constant-strain minimization method is employed to the available equilibrated system. To obtain the stress–strain relationship, a tensile deformation is applied to the systems by increasing the box length in the tensile direction and remapping the atom coordinates in every step. To allow the natural Poisson contraction stresses perpendicular to the tensile direction are set to be zero. 

\begin{figure}[h!]
	\centering
	\includegraphics[width=0.5\textwidth]{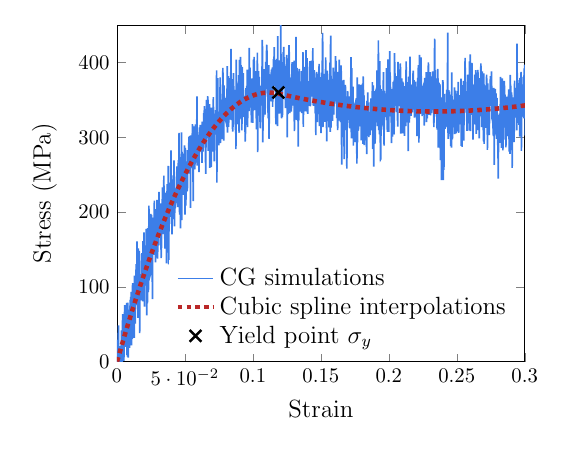}
	\caption{Stress–strain relationship of the pure epoxy obtained from CG simulations. Simulation data points have been averaged over tensile simulations of five different configurations.} 
	\label{knot}
\end{figure} 

To obtain an average stress–strain response, tensile simulations are repeated in the X, Y, and Z directions for five different configurations. The tensile simulations are performed at different strain rates and temperatures below the glass transition point. The time step is set to be 5 fs. 

Fig.~\ref{knot} shows a representative stress–strain curve of a cured epoxy system at strain rate $\textstyle \dot{\epsilon}$ = $\textstyle 10^{8}$~1/s. A piecewise cubic spline interpolation with an optimized knot is fitted to the simulation data points. The position of each knot is optimized by minimizing the least square error between the fit and data points. The maximum of the fitted curve marked by a cross in the figure is taken as the yield point.

To eliminate the effects of the boundary in small molecular systems, periodic boundary conditions are used to calculate the material properties. Although  periodic boundary conditions allow removing the artifact caused by unwanted boundaries, the effect of unwanted boundaries is replaced by the artifact of periodic conditions. Therefore, considering the influence of periodic boundary conditions on simulation predictions is necessary. For amorphous polymer systems, the motion of one molecule intensely affects neighbouring molecules. The application of periodic conditions leads to underestimated predictions compared to those in infinite systems \cite{dunweg1991microscopic}. Therefore, a sufficiently large sample volume of the polymer materials has to be selected. 

\begin{figure}[h!]
	\centering
	\includegraphics[width=0.5\textwidth]{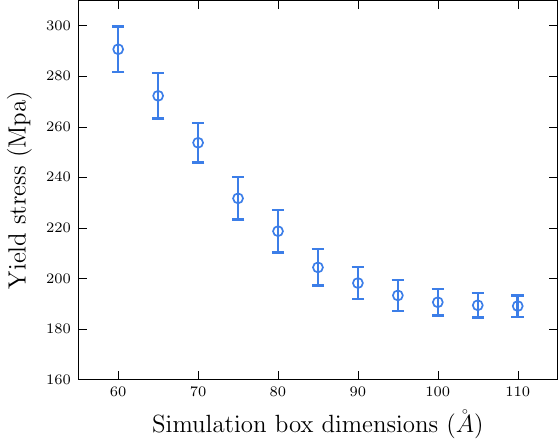}
	\caption{Variation of $\sigma_{y}$ of a pure epoxy system with respect to the simulation box side length at a strain rate of $\textstyle \dot{\epsilon}$ = $\textstyle 10^{8}$~1/s.} 
	\label{polymerSize00}
\end{figure} 

To choose a suitable representative volume element size for the epoxy system, periodic simulation boxes with side lengths varying from 60 to 110~$\text{\AA}$ are simulated. 

The uncured system is initially subjected to an energy minimization to find a global minimum energy configuration of the system. The curing process and preparing is performed as explained in subsection \ref{subsection:CGSim}. To extract the stress–strain curve, the systems are then subjected to uniaxial loading at the strain rate of $\textstyle \dot{\epsilon}$ = $\textstyle 10^{8}$~1/s and room temperature. Next, the average yield stress is calculated as shown in Fig.~\ref{polymerSize00}. Each data point is obtained from simulations of five different configurations. The simulation results show a decrease in the average yield stress by increasing the side lengths from 60 to 95~$\text{\AA}$. 
Based on Fig.~\ref{polymerSize00}, simulation results show an asymptotic behavior for side lengths larger than 100~$\text{\AA}$. Therefore, a periodic box with a side length larger than 100~$\text{\AA}$ can represent the polymer structure.

\subsection{Parameter identification of the epoxy}
\label{subsection:Constitutive_epoxy}

The unknown parameters of the constitutive model consist of ten different values for the nonlinear dashpot ($\dot{\epsilon}_{0}$, $\sigma_{0}$, $\Delta H $), hyperelastic spring ($\mu_{he,ref}$, $\alpha_{he}$, $N$), linear elastic spring ($\mu_{e,ref}$, $\lambda_{e,ref}$, $\alpha_{e}$), and damage parameter ($A$). 

\begin{figure}[h!]
	\centering
	\begin{subfigure}[b]{0.45\linewidth}
		\centering
		\includegraphics[width=\textwidth]{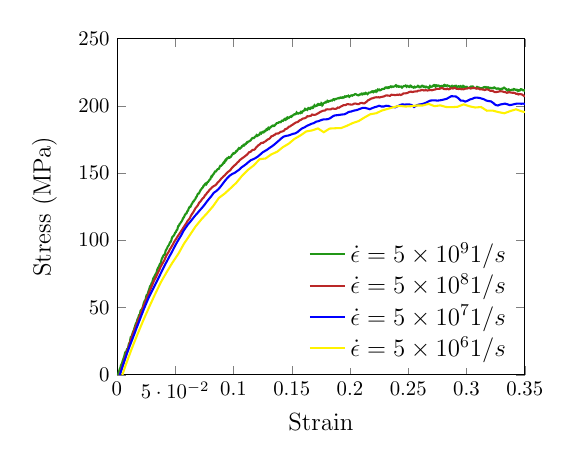}
		\caption{}
		\label{stress_strain_temp1}
	\end{subfigure}
	\begin{subfigure}[b]{0.45\linewidth}
		\centering
		\includegraphics[width=\textwidth]{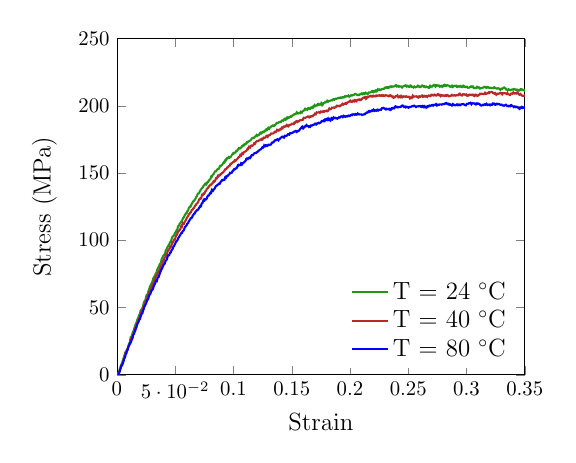}
		\caption{}
		\label{stress_strain_temp2}  
	\end{subfigure}    
	\caption{Stress–strain response obtained by CG simulations (a) at various strain rates at room temperature and (b) three different temperatures (24, 40 and 80~$^{\circ}$C) at a constant strain rate of  $\textstyle \dot{\epsilon}$ = $\textstyle 5\times 10^{9}$~1/s.} 
	\label{stress_strain_temp}
\end{figure}

To determine the parameters of the viscoelastic dashpot, the Argon model presented in Eq. (12) is parameterized using CG tensile simulations. It is worth noting that CG simulations are performed at strain rates several orders of magnitude higher than those applied in experiments which is a well-known limitation of molecular simulations due to computational constraints. 
The Argon model bridges this gap by establishing a physically based relationship between yield stress, strain rate, and temperature, allowing parameters identified at CG strain rates to be extrapolated to experimentally relevant rates. The validity of this extrapolation rests on the assumption that the thermally activated flow mechanism governing yielding remains the same across the entire rate range. The assumption is supported by prior work on epoxy systems of similar chemistry, where Argon-based extrapolations from MD-accessible rates to experimental rates have been shown to yield physically consistent parameters \cite{arash2019viscoelastic, unger2019non}.
The yield stress is obtained as a function of the strain rate and temperature as shown in Figs.~\ref{stress_strain_temp}. To study the effect of strain rate and temperature on the stress–strain behavior of the pure epoxy, CG tensile simulations are performed at different strain rates ranging from $\textstyle \dot{\epsilon}$ = $\textstyle 5\times 10^{6}$ to $\textstyle \dot{\epsilon}$ = $\textstyle 5\times 10^{9}$~1/s at three different temperatures of 24, 40 and 80~$^{\circ}$C. The temperatures are chosen based on the temperatures applied at the experimental tensile tests to cover the glassy regime. To ensure a well-equilibrated system, the system is initially relaxed at the simulation temperature for 20~ns. 

\begin{figure}[h!]
	\centering
	\begin{subfigure}[b]{0.3\linewidth}
		\centering
		\includegraphics[width=\textwidth]{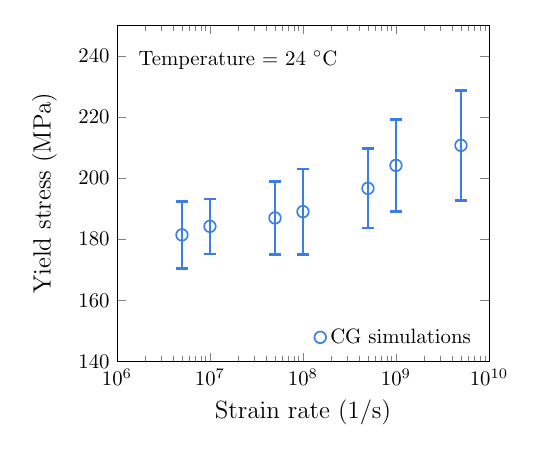}
		\caption{}
		\label{ST_temps1}
	\end{subfigure}
	\begin{subfigure}[b]{0.3\linewidth}
		\centering
		\includegraphics[width=\textwidth]{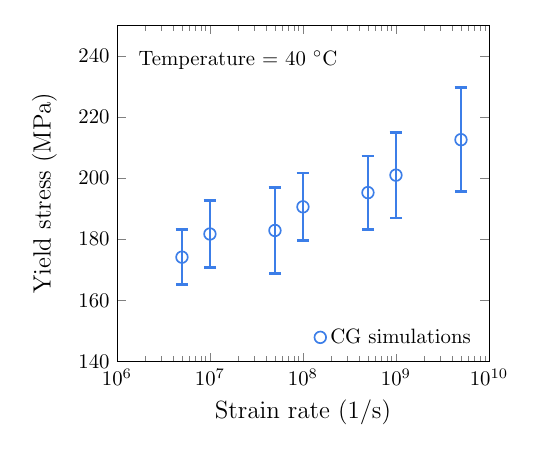}
		\caption{}
		\label{ST_temps2}  
	\end{subfigure} 
	\begin{subfigure}[b]{0.3\linewidth}
		\centering
		\includegraphics[width=\textwidth]{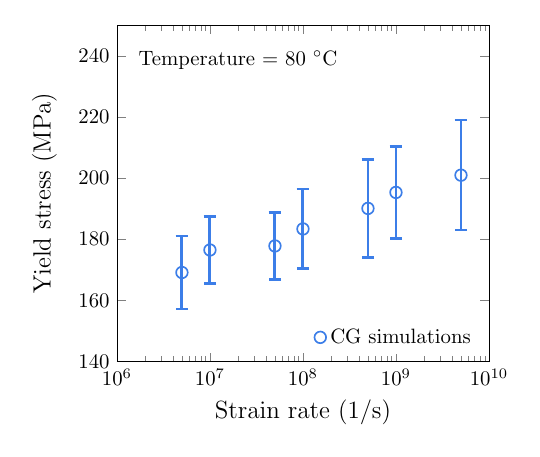}
		\caption{}
		\label{ST_temps3}  
	\end{subfigure}        
	\caption{Yield stress as a function of strain rate obtained from CG simulations at three different temperatures (all below the glass-transition temperature).} 
	\label{ST_temps}
\end{figure}

To identify the material behavior in the nonlinear regime, the maximum stress is obtained at which viscous flow occurs. To find the yield stress in stress–strain curves, cubic spline interpolation is used as described in the previous section. Fig.~\ref{stress_strain_temp1} shows the averaged stress–strain curves for strain rates ranging from $\textstyle \dot{\epsilon}$ = $\textstyle 5\times 10^{6}$~1/s to $\textstyle \dot{\epsilon}$ = $\textstyle 5\times 10^{9}$~1/s at the room temperature. The stress–strain responses of CG simulations for pure epoxy show that the yield stress increases by increasing the strain rate. Fig.~\ref{stress_strain_temp2} shows the average stress–strain curve of the cured epoxy system at three different temperature (24, 40 and 80~$^{\circ}$C) obtained by CG simulations at a constant strain rate of  $\textstyle \dot{\epsilon}$ = $\textstyle 5\times 10^{9}$~1/s. As a result, the yield stress for a constant strain rate decreases by increasing temperature. It is worth noting that the largest standard deviation of all the averaged results has a value of around 18 MPa which is in the typical scattering range of molecular simulations.
The stress-strain curves in Figs.~\ref{stress_strain_temp} show relatively small differences across the range of strain rates and temperatures, which are comparable to the simulation scatter. However, the yield stress extracted from these curves shows a clear and systematic dependence on both strain rate and temperature, as shown in Fig.~\ref{ST_temps}. This confirms that the trends are physically consistent and sufficient to identify the parameters of the Argon model.

Having the simulation results for the yield stress at different strain rates and temperature, an in-house grid search algorithm is used to obtain an optimized set of parameters for the Argon model. In this approach, the objective function is evaluated at uniformly distributed points within the parameter space, and the search boundaries are iteratively decreased around the optimal solution until convergence is achieved.
To determine the size of the search space, some parametric studies were carried out. The objective function is calculated by a deviation between predicted yield stress values and yield stress obtained by CG simulation results. For each combination of strain rate and temperature, we can obtain mean values from CG simulations and use for the parameter identifications. Accordingly, the optimized parameters of the viscoelastic dashpot are extracted to be $\dot{\epsilon}_{0}$ = 1.1379$\times10^{12}$ 1/s, $\sigma_{0}$ = 235.9785 MPa and $\Delta H $ = 1.9843$\times$$10^{-19}$ J. The identified parameters are in fair agreement with existing values in the literature by other scientists for similar epoxy systems \cite{unger2019non,arash2019viscoelastic,arash2021finite}.

According to the cross-linking simulation, the parameter $N$, indicating the average number of rigid links between two cross-links, is obtained to be 2.44. Also, we assume that the temperature scaling parameters are equal, namely $\alpha_{he}=\alpha_{e}$. This assumption is motivated by the observation that both the rubbery network stiffness and the glassy modulus of crosslinked epoxies are governed by the same segmental mobility and that their temperature dependences track each other closely below the glass transition \cite{unger2019non}.
Therefore, the number of remaining unknown parameters reduces to four (i.e., $\mu_{he,ref}$, $\mu_{e,ref}$, $\lambda_{e,ref}$, $\alpha_{e}$). To identify the remaining parameters of the constitutive model, a single-objective genetic algorithm \cite{golberg1989genetic} of an in-house optimization framework was used. The stress–strain curve of five individual experimental results, including two strain rates and three different temperatures, is used in the optimization. For each stress–strain curve, the error was defined by the difference between the experimental data and material model prediction. The objective function was defined to be the euclidean norm of all five individual errors. Several optimization runs were performed, of which all converged to a minimum. The final identified parameters are presented in Table~\ref{cons_epoxy}. 

\begin{table}[h!]
	\caption{Identified parameters of the constitutive model of pure epoxy using simulation and experimental data. Superscripts $^{A}$ and $^{E}$ denote the Argon viscoelastic model and epoxy, respectively.}
	\label{cons_epoxy}
	\centering
	\begin{tabular}	{  l     l     l  }
		\toprule
		\multicolumn{1}{C{5cm}}{{\normalsize }} & 
		\multicolumn{1}{C{3cm}}{{\normalsize Parameter }} & 
		\multicolumn{1}{C{3cm}}{{\normalsize Value}} \\
		
		\cmidrule(r){1-3}
        
		Hyperelastic spring & $\mu_{he,ref}$ (MPa)  & 396.5795\\
		& $\alpha_{he}$  & $0.00409$\\
		& $N$ & $2.44$\\
		\bottomrule

		Linear spring &  $\mu_{e,ref}$ (MPa)  & 841.8711\\
		& $\lambda_{e,ref}$ (MPa) & $3367$\\
		& $\alpha_{e}$ & $0.00409$\\
		\bottomrule
		
		Nonlinear dashpot &  $\sigma_{0}^{A}$ (MPa) & 235.9785\\
		& $\Delta H^{E}$ (J) & 1.9843$\times$$10^{-19}$\\
		& $\dot{\epsilon}_{0}^{E}$ (1/s) & 1.1379$\times10^{12}$\\
		\bottomrule
		
		Damage &  $A$ & 166.2944\\
		\bottomrule
		
	\end{tabular}
\end{table}

\subsection{Parameter identification of nanoparticle effects}
\label{subsection:Constitutive_nano}

BNPs are nonuniformly dispersed in an epoxy matrix \cite{khorasani2019effect}, mainly forming agglomerates and rarely occurring as single particles. It implies that nanoparticle distribution and weight fraction significantly affect the nanocomposites' material behavior. In the following subsection, CG simulations are performed to capture the effect of the two parameters on the nanocomposites' stress–strain relationship. Simulation results are used to calibrate the amplification factor presented in Eq. \ref{Constitutive5} as a function of the nanoparticle weight fraction and agglomerate size. The amplification factor allows considering the effect of BNPs in the elastic and hyperelastic spring elements (see Fig.~\ref{rheological_model}). Furthermore, the large-scale CG simulations enable us to study the viscoelastic flow of nanoparticle-filled epoxies. For this, the parameters of the Argon viscoelastic model are extracted as a function of the BNP weight fraction and agglomerate size.

\begin{figure}[h!]
	\centering
	\begin{minipage}{0.75\textwidth}  
		\centering
		\begin{subfigure}[c]{0.15\textwidth}
			\centering
			\includegraphics[width=\textwidth]{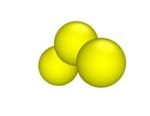}
		\end{subfigure}
		\hfill
		$\Large\cdots$
		\hfill
		\begin{subfigure}[c]{0.28\textwidth}
			\centering
			\includegraphics[width=\textwidth]{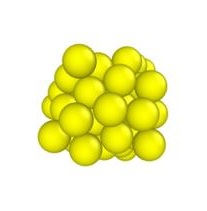}
		\end{subfigure}
		\hfill
		$\Large\cdots$
		\hfill
		\begin{subfigure}[c]{0.38\textwidth}
			\centering
			\includegraphics[width=\textwidth]{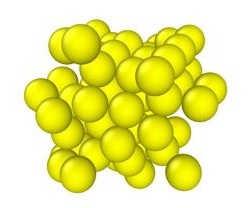}
		\end{subfigure}
		
		\vspace{1.0em}
		\noindent\hspace*{-0.1\textwidth}%
		\begin{tikzpicture}
			\draw[thick, -stealth, line width=1.2pt]
			(0,0) -- (0.735\textwidth, 0);
			\node[above=5pt, anchor=west] at (0,0) 
			{\small 2.3~nm};
			\node[above=5pt, anchor=east] at (0.735\textwidth, 0) 
			{\small 10~nm};
		\end{tikzpicture}
	\end{minipage}
	
	\caption{BNP agglomerates with 3, 40, and 80 primary 
		particles, covering agglomerate sizes ranging from 2.3~nm to 10~nm.}
	\label{aggSizes}
\end{figure}
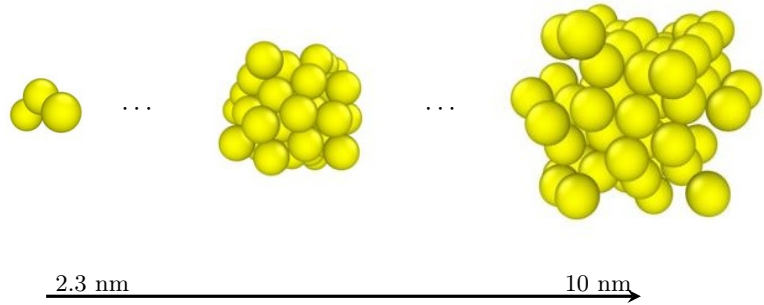
\begin{figure}[h!]
	\centering
	\begin{subfigure}[b]{0.4\linewidth}
		\centering
		\includegraphics[width=\textwidth]{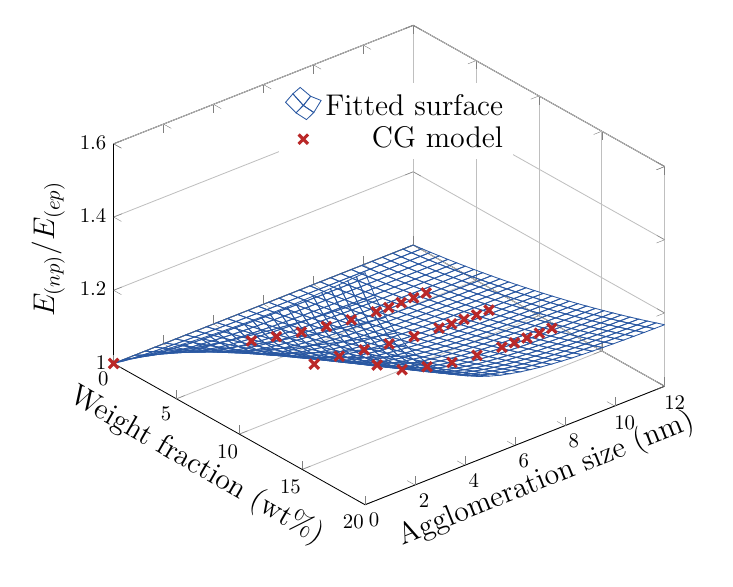}
		\caption{}
		\label{3d_E1}  
	\end{subfigure}  
	\begin{subfigure}[b]{0.29\linewidth}
		\centering
		\includegraphics[width=\textwidth]{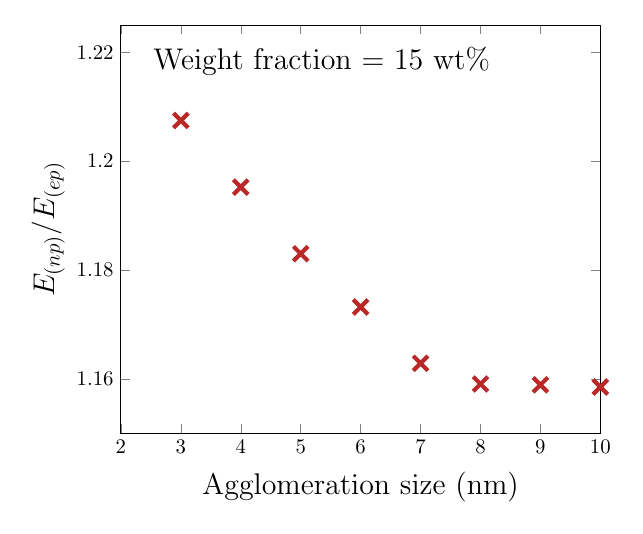}
		\caption{}
		\label{3d_E2}  
	\end{subfigure}    
	\begin{subfigure}[b]{0.29\linewidth}
		\centering
		\includegraphics[width=\textwidth]{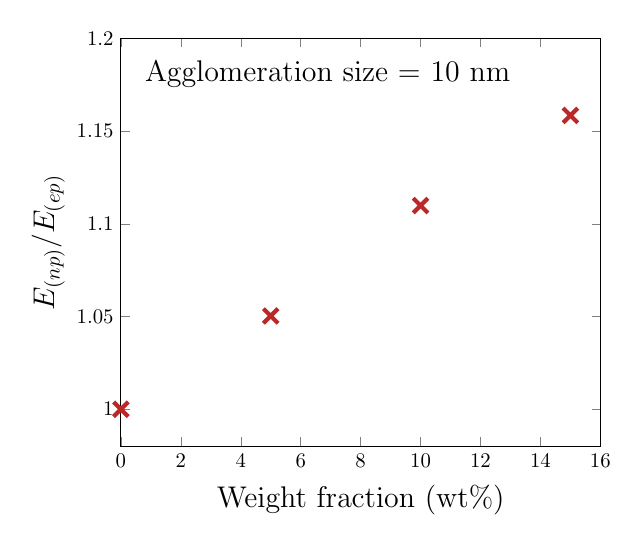}
		\caption{}
		\label{3d_E3}  
	\end{subfigure} 
	\caption{(a) Variation ratio of the effective Young’s modulus of the BNP/epoxy nanocomposites with respect to the nanoparticle weight fraction and agglomerate size, (b) the effect of the agglomorate size on the Young’s modulus of the nanocomposites, and (c) the effect of the BNP weight fraction on the Young’s modulus of the nanocomposites.} 
	\label{3d_E}
\end{figure}

To investigate the effect of the agglomerate size on the stiffness, tensile simulations of agglomerated BNP/epoxy nanocomposites are performed as described in the previous section. For this purpose, agglomerate sizes with the average diameter varying from 23~$\text{\AA}$ to 100~$\text{\AA}$ are constructed. To increase agglomerate sizes, the number of BNPs increases from 3 to 80, as shown in Fig.~\ref{aggSizes}. For each agglomerate size, three cured BNP/epoxy models with weight fractions of 5, 10, and 15 \%wt subjected to uniaxial loading are simulated. Therefore, each data point in Fig.~\ref{3d_E} represents the average Young's modulus calculated for three different configurations along the x-, y-, and z-axis. 

\begin{figure}[h!]
	\centering
	\begin{subfigure}[b]{0.30\linewidth}
		\centering
		\includegraphics[width=\textwidth]{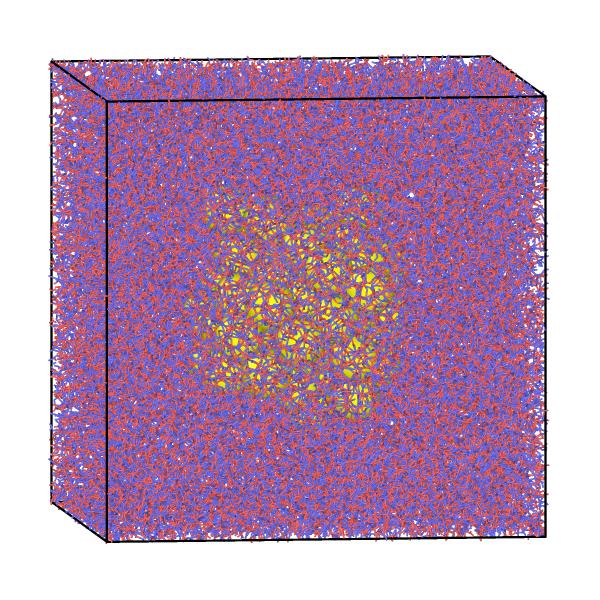}
		\caption{}
		\label{box1}  
	\end{subfigure}    
	\begin{subfigure}[b]{0.30\linewidth}
		\centering
		\includegraphics[width=\textwidth]{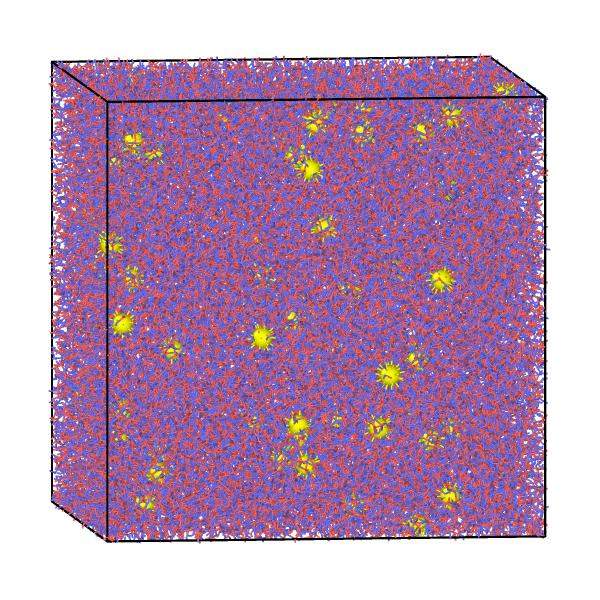}
		\caption{}
		\label{box2}  
	\end{subfigure}    
	\caption{A cured simulation box of CG model with a size of 280$\times$280$\times$280~$\text{\AA}$\textsuperscript{3} for (a) fully agglomerated BNP(5\%)/epoxy nanocomposites (BNP size of 100~$\text{\AA}$) and (b) well dispersed BNP(5\%)/epoxy nanocomposites  (BNP size of 23~$\text{\AA}$).}
	\label{Boxes}
\end{figure}

The simulation box of BNP(5\%)/epoxy nanocomposites with well dispersed and fully agglomerated BNPs with the agglomerate size of 100~$\text{\AA}$ (10~nm) is shown in Figs.~\ref{Boxes}. The number of beads is respectively 70034 and 62774, which corresponds to 2211088, and 2031352 atoms in all-atom models, respectively.

Fig.~\ref{3d_E1} presents the variation of the effective Young’s modulus of the nanocomposites to Young’s modulus pure epoxy ratio ($E(np)/E(ep)$) with respect to the BNP weight fraction and agglomerate size. As can be seen in Fig.~\ref{3d_E2}, the stiffness ratio converges to ~1.16 by increasing the agglomeration size to 100~$\text{\AA}$, while the weight fraction kept constant and equal to 15~wt\%. Fig.~\ref{3d_E3} also shows that the ratio linearly increases by increasing the BNP weight fraction, while the agglomeration size is kept constant and equal to 100~$\text{\AA}$. To consider the effects of agglomeration size together with nanoparticle weight fraction, a closed-form equation for the amplification factor $X$ is suggested as follows:

\begin{equation}
\begin{split}
X = a W  e^{-b S} + c W^{2} + d W + 1,   
\end{split}
\label{amplification_formula}
\end{equation}
where $S$ is the agglomeration size, and $W$ is the BNP weight fraction. The equation is suggested by fitting a surface to the simulation data points as shown in Fig.~\ref{3d_E1}. The optimized values for $a$, $b$, $c$ and $d$ are 0.0234, 0.4448, 0.000195 and 0.0044, respectively.

All the models prepared for the previous subsection are used here to investigate the effects of nanoparticles on the viscoelastic flow. For each set of agglomerate size and weight fraction, the parameters of the Argon model presented in Eq. \ref{Constitutive12} (i.e., the pre-exponential factor $\dot{\epsilon}_{0}$, athermal yield stress ${\sigma}_{0}$, and activation energy ${\Delta}H$) are extracted using the simulation method described in Section~\ref{subsection:Constitutive_epoxy}. 

\begin{figure}[h!]
	\centering
	\begin{subfigure}[b]{0.35\linewidth}
		\centering
		\includegraphics[width=\textwidth]{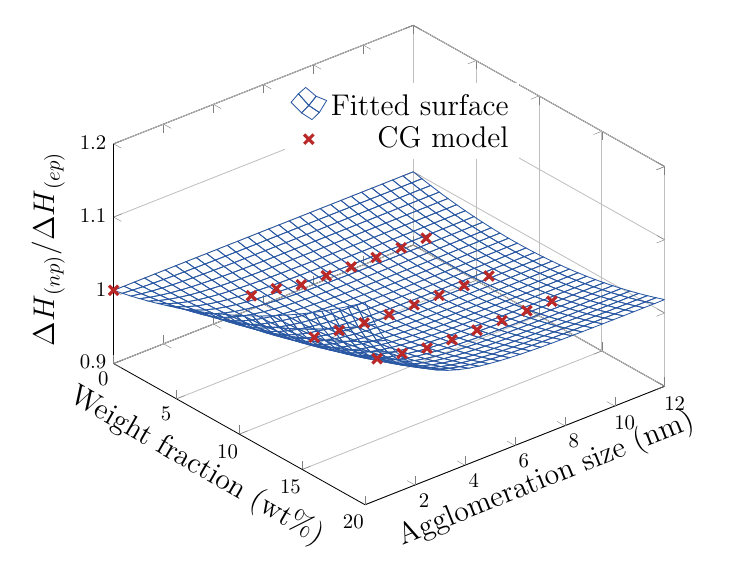} 
		\caption{}
		\label{H1}  
	\end{subfigure}   
	\begin{subfigure}[b]{0.3\linewidth}
		\centering
		\includegraphics[width=\textwidth]{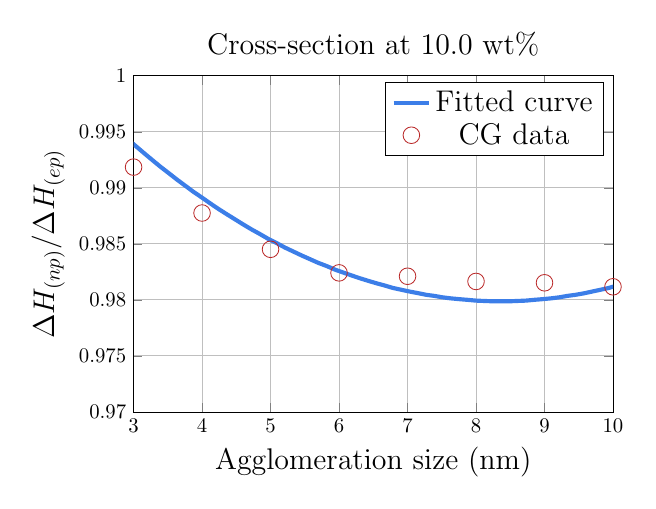} 
		\caption{}
		\label{H3}  
	\end{subfigure}   
	\begin{subfigure}[b]{0.3\linewidth} 
		\centering
		\includegraphics[width=\textwidth]{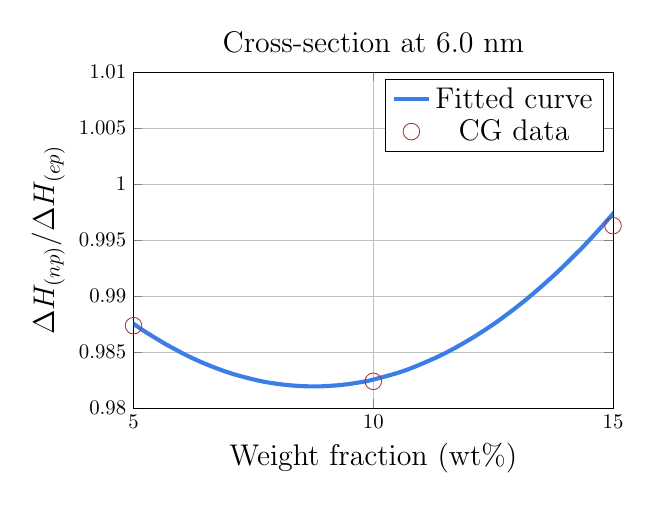} 
		\caption{}
		\label{H4}  
	\end{subfigure} 
	\caption{Detailed variation of the activation energy ratio ${\Delta}H_{(np)}$/${\Delta}H_{(ep)}$ of the Argon viscoelastic model with respect to the BNP weight fraction and agglomerate size: (a) fitted 3D surface, (b) cross-section at a constant weight fraction of 10~wt\%, and (c) cross-section at a constant agglomerate size of 6~nm.} 
	\label{H}
\end{figure} 

\begin{figure}[h!]
	\centering
	\begin{subfigure}[b]{0.35\linewidth}
		\centering
		\includegraphics[width=\textwidth]{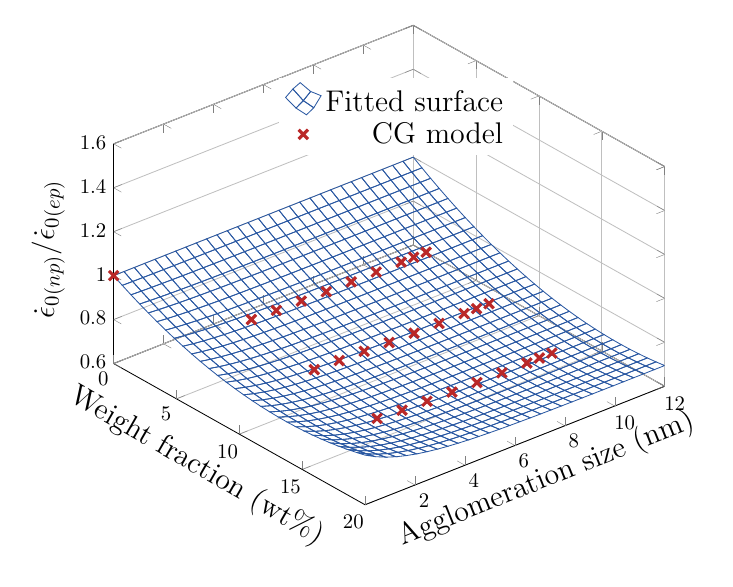} 
		\caption{}
		\label{eps1}  
	\end{subfigure}   
	\begin{subfigure}[b]{0.3\linewidth}
		\centering
		\includegraphics[width=\textwidth]{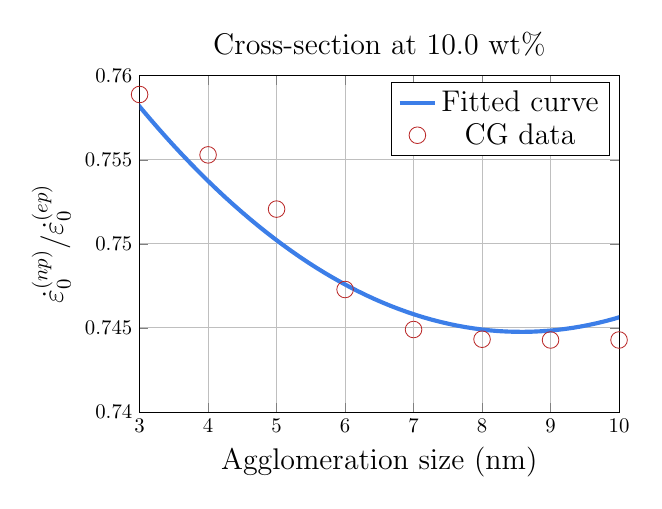} 
		\caption{}
		\label{eps3}  
	\end{subfigure}   
	\begin{subfigure}[b]{0.3\linewidth}
		\centering
		\includegraphics[width=\textwidth]{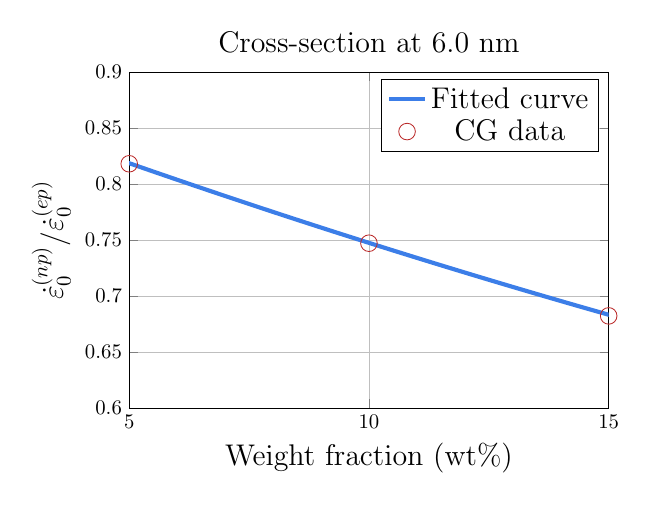}
		\caption{}
		\label{eps4}  
	\end{subfigure} 
	\caption{Detailed variation  of the pre-exponential factor ratio of $\dot{\varepsilon}_0^{(np)}/\dot{\varepsilon}_0^{(ep)}$ the Argon viscoelastic model with respect to the BNP weight fraction and agglomerate size: (a) fitted 3D surface, (b) cross-section at a constant weight fraction of 10~wt\%, and (c) cross-section at a constant agglomerate size of 6~nm.}
	\label{eps}
\end{figure}

\begin{figure}[h!]
	\centering
	\begin{subfigure}[b]{0.35\linewidth}
		\centering
		\includegraphics[width=\textwidth]{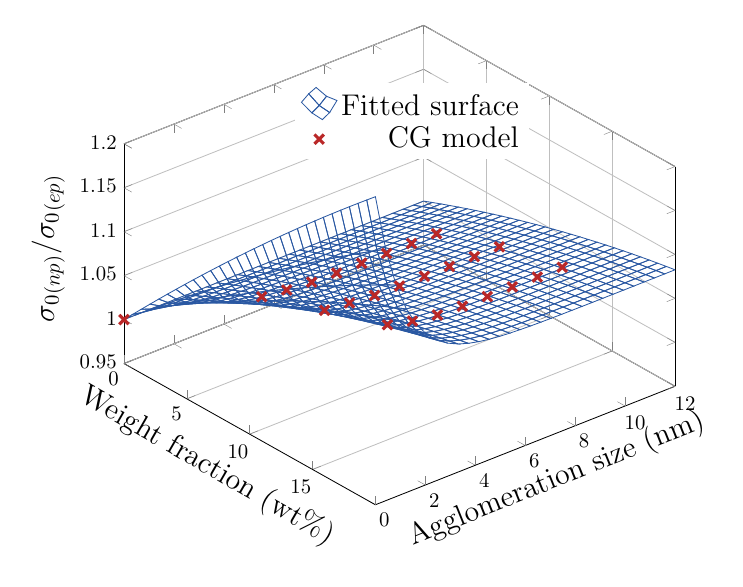} 
		\caption{}
		\label{sigma1}  
	\end{subfigure}   
	\begin{subfigure}[b]{0.3\linewidth}
		\centering
		\includegraphics[width=\textwidth]{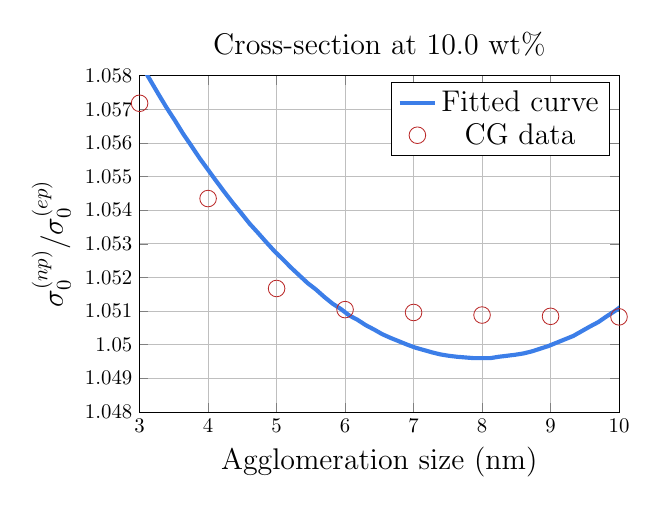} 
		\caption{}
		\label{sigma3}  
	\end{subfigure}   
	\begin{subfigure}[b]{0.3\linewidth}
		\centering
		\includegraphics[width=\textwidth]{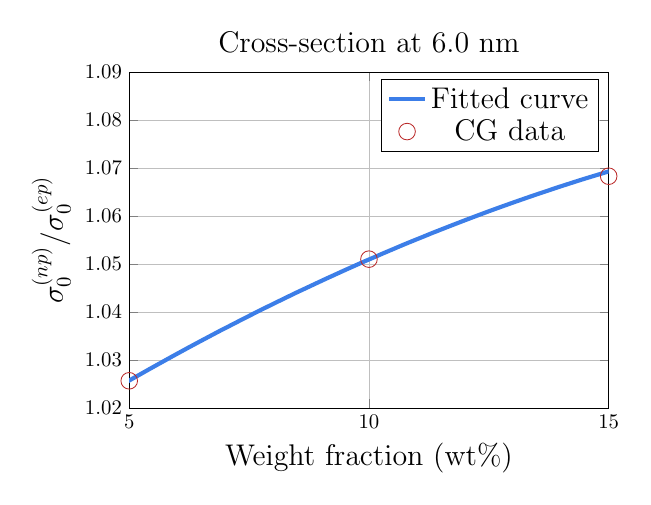}
		\caption{}
		\label{sigma4}  
	\end{subfigure} 
	\caption{Detailed variation of the athermal yield stress $\sigma_{0}^{(np)}/\sigma_{0}^{(ep)}$ of the Argon viscoelastic model with respect to the BNP weight fraction and agglomerate size: (a) fitted 3D surface, (b) cross-section at a constant weight fraction of 10~wt\%, and (c) cross-section at a constant agglomerate size of 6~nm.} 
	\label{sigma}
\end{figure}

Here, $np$, and $ep$ subindices represent a parameter for the nanocomposites and pure epoxy. Figs.~\ref{H}-\ref{sigma} show the variation of $\dot{\epsilon}_{0}(np) / \dot{\epsilon}_{0}(ep) $, ${\sigma}_{0}(np) / {\sigma}_{0}(ep)$, and ${\Delta}H_{(np)} / {\Delta}H_{(ep)}$ ratios with respect to the BNP agglomerate size and weight fraction, respectively. Each data point represents the average amount of the parameters obtained from the simulation of three different configurations. Similar to the amplification factor, the following closed-form equation, showing the variation of the ratios in the presence of BNPs, is suggested for each of the parameters. 

\begin{equation}
	\begin{split}
		f = a W  e^{-b S} + c W^{2} + d W + 1,   
	\end{split}
	\label{aron_formula}
\end{equation}

where $S$ is the agglomeration size, and $W$ is the BNP weight fraction. For each parameter of the Argon model, the equation unknowns (i.e., $a$, $b$, $c$ and $d$ ) are optimized by fitting a surface to the simulation data points as shown in Figs.~\ref{H}-\ref{sigma}. The optimized values are listed in Table~\ref{amplication_factor}. 

The fitted surfaces obtained from Eq.~\ref{aron_formula} show good agreement with the CG simulation data points across the entire parameter space, confirming the suitability of the proposed closed-form expression. As seen in Fig.~\ref{eps}, the pre-exponential factor ratio $\dot{\varepsilon}_0^{(np)}/\dot{\varepsilon}_0^{(ep)}$ decreases with increasing BNP weight fraction, indicating that the presence of nanoparticles reduces the characteristic viscoelastic flow rate of the nanocomposite. This behavior is physically consistent with the notion that nanoparticles act as obstacles to polymer chain mobility, increasing the resistance to viscous flow. In contrast, Fig.~\ref{sigma} shows that the athermal yield stress ratio $\sigma_{0}^{(np)}/\sigma_{0}^{(ep)}$ increases with BNP content, which reflects the mechanical reinforcement of the matrix and the increased stress required to initiate irreversible deformation. The activation energy ratio ${\Delta}H_{(np)}$/${\Delta}H_{(ep)}$, shown in Fig.~\ref{H3}, remains close to unity across the range of weight fractions and agglomerate sizes considered, suggesting that the fundamental energy barrier governing viscoelastic flow is not significantly altered by the presence of BNPs. The relatively weak dependence of ${\Delta}H$ on nanoparticle content is in agreement with findings reported for similar nanoparticle/epoxy systems in the literature \cite{arash2019viscoelastic, unger2019non}. 

Furthermore, the effect of agglomerate size on all three parameters is moderate compared to the effect of weight fraction, which is consistent with the observation that the amplification factor X is primarily governed by BNP content rather than by the degree of agglomeration. Collectively, these results demonstrate that the proposed CG simulation-based framework successfully captures the complex dependence of viscoelastic flow on nanoparticle morphology, providing a physically meaningful and computationally efficient basis for the constitutive model.

\begin{table}[h!]
	\caption{Optimized parameters of Eq.~\ref{aron_formula} for the Argon viscoelastic model.}
	\label{amplication_factor}
	\centering
	\begin{tabular}	{  c     c     c   c  c}
		\toprule
		\multicolumn{1}{C{3cm}}{{\normalsize Argon model}} & 
		\multicolumn{1}{C{2cm}}{{\normalsize a }} & 
		\multicolumn{1}{C{2cm}}{{\normalsize b }} & 
		\multicolumn{1}{C{2cm}}{{\normalsize c }} &
		\multicolumn{1}{C{2cm}}{{\normalsize d }} \\ 
		\cmidrule(r){1-5}
		
		$\dot{\epsilon}_{0}$& 0.0066  & 0.5247 & 0.0012 & -0.0393\\
		${\sigma}_{0}$ 	 & 0.0109  &  0.8454 & -0.00008375 & 0.0058\\
		${\Delta}H$ &  0.0079 &   0.6151 &  0.00026057 & -0.0043\\
		
		\bottomrule
		
	\end{tabular}
\end{table} 

Using the effects of nanoparticles captured by Eqs.~\ref{amplification_formula} and \ref{aron_formula} and the identified parameters for the pure epoxy listed in Table~\ref{cons_epoxy}, the proposed constitutive model can predict the stress–strain relationship of the BNP/epoxy nanocomposites at different strain rates and temperatures. In the next subsection, we evaluate the predictive capability of the constitutive model.

\subsection{Experimental validation}
\label{subsection:validation}

Experimental tensile tests are performed according to the test procedure explained in Section~\ref{section:ExP_explain}. The tests are conducted at two strain rates of $\textstyle \dot{\epsilon} = \textstyle 1.67\times 10^{-5}$ and $\textstyle 1.67\times 10^{-4}$~1/s and three different temperatures of 24, 40 and 80~$^{\circ}$C for pure epoxy, BNP(10 wt\%)/epoxy and BNP(15 wt\%)/epoxy. Each experimental data point is the mean value of ten measurements. 

\begin{figure}[h!]
	\centering
	\includegraphics[width=0.6\textwidth]{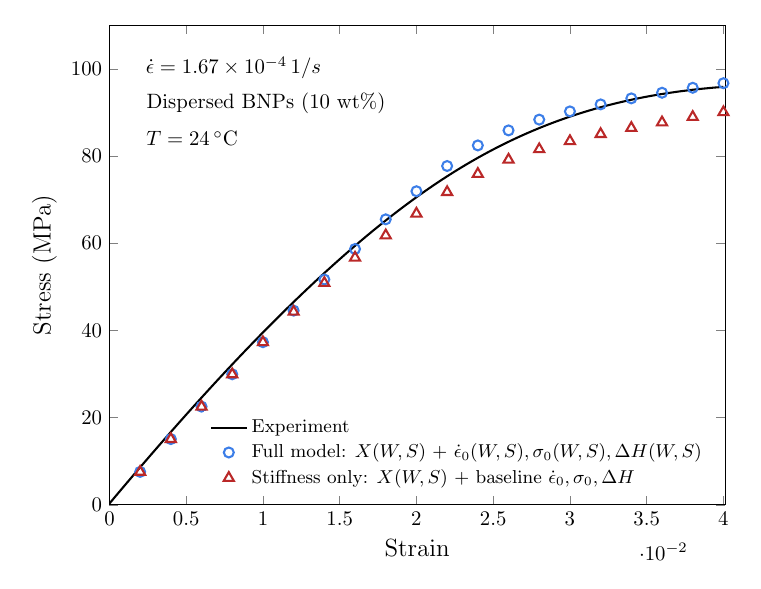}
	\caption{Stress--strain response of dispersed BNP/epoxy nanocomposites
		(10~wt\%) predicted by the full model and by a stiffness-only variant in
		which the amplification factor $X(W,S)$ is retained but the Argon flow
		parameters are held at their pure-epoxy baseline values, compared with
		experiment.}
	\label{fig:amplification_ablation}
\end{figure} 

Before assessing the model across the full range of conditions, we first isolate
the distinct roles played by the two morphology-dependent mechanisms of the
constitutive model. Figure~\ref{fig:amplification_ablation} compares the
experimental response of the 10~wt\% nanocomposite with two model variants: the
full model, in which both the amplification factor $X(W,S)$ and the Argon flow
parameters $\dot{\epsilon}_0(W,S)$, $\sigma_0(W,S)$ and $\Delta H(W,S)$ depend on
morphology, and a stiffness-only variant, in which $X(W,S)$ is retained but the
flow parameters are held at their pure-epoxy baseline values. The two variants
are nearly indistinguishable in the initial elastic regime, since the Argon
parameters govern only the rate of inelastic flow and therefore become active
once yielding begins. Beyond yield, the stiffness-only variant progressively
underestimates the measured stress, reaching a deviation of roughly $6$~MPa at
$\varepsilon=0.04$, whereas the full model remains in close agreement throughout.
Quantitatively, the full model reproduces the experimental curve with
$R^2=0.997$ (RMSE${}=1.6$~MPa), while removing the morphology dependence from the
flow parameters reduces the agreement to $R^2=0.977$ (RMSE${}=4.2$~MPa). This
demonstrates that the amplification factor and the morphology-dependent flow
kinetics act in different deformation regimes and contribute independently to the
reinforcement: the former controls the elastic and hyperelastic stiffening, while
the latter governs the post-yield viscoelastic response. The two mechanisms are
therefore complementary rather than redundant, confirming that scaling the
non-equilibrium stress by $X(W,S)$ does not double-count the morphology effects
already embedded in the Argon flow rule.

\begin{figure}[h!]
	\centering
	\begin{subfigure}[b]{0.48\linewidth}
		\centering
		\includegraphics[width=\textwidth]{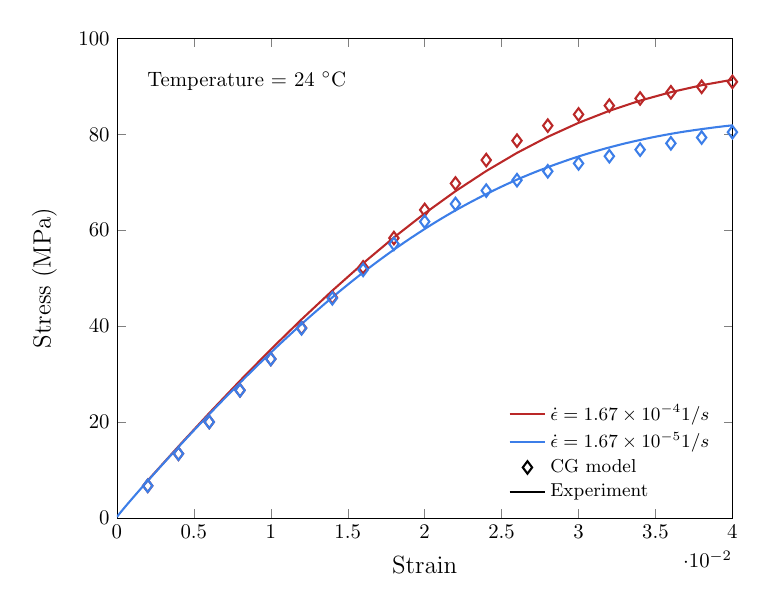}
		\caption{}
		\label{visco_epoxy1}
	\end{subfigure}
	\begin{subfigure}[b]{0.48\linewidth}
		\centering
		\includegraphics[width=\textwidth]{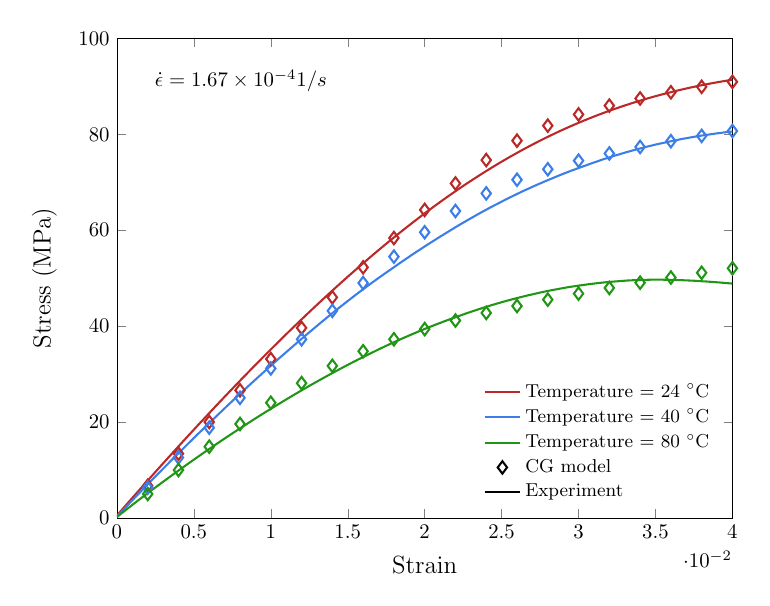}
		\caption{}
		\label{visco_epoxy2}  
	\end{subfigure}    
	\caption{Effect of strain rate (a) and temperature (b) on the stress–strain relationship of the epoxy system.} 
	\label{visco_epoxy}
\end{figure}

Figs.~\ref{visco_epoxy1} shows the stress–strain relationship of the pure epoxy at the strain rate of $\textstyle 1.67\times 10^{-4}$ and $\textstyle 1.67\times 10^{-5}$~1/s. Also, to evaluate the predictive capability of the CG simulation-informed constitutive model, the predicted stress–strain curves are compared with experimental data at three temperatures of 24, 40 and 80~$^{\circ}$C in Fig.~\ref{visco_epoxy2}. Overall, the predicted stress–strain curves show satisfactory agreement with the experimental data, capturing both the initial linear response and the nonlinear softening behavior. From Figs.~\ref{visco_epoxy2}, the most significant deviation is observed at the strain rate of $\textstyle 1.67\times 10^{-4}$~1/s at 80 $^{\circ}$C, where the yield point occurs earlier compared to the experimental results. Although the initial stiffness and post-peak softening behavior are well predicted for all configurations, the prediction accuracy is reduced at 80~$^{\circ}$C due to the increasing sensitivity of the Kitagawa exponential scaling at larger temperature offsets from reference and the interpolation uncertainty in the CG-derived Argon model parameters near the upper boundary of the calibration range.
Furthermore, the effect of strain rate on the stress–strain relationship is accurately captured by the calibrated constitutive model. It is worth mentioning that the material response in this regime is dominated by the hyperelastic and viscoelastic contributions rather than damage-induced softening. The damage mechanism becomes significant only at larger strains approaching failure, which were not the focus of the present validation.

It is worth mentioning that the distribution of BNPs in the epoxy matrix is not fully known. Therefore, in the following experimental-numerical comparisons, the amplification factor and the parameters of the Argon model of the constitutive model are calculated for two extreme cases, namely fully distributed and agglomerated BNPs with the largest diameter simulated in Sec. \ref{subsection:Constitutive_nano} (see Fig. \ref{Boxes}). Accordingly, a lower and upper bounds of the stress–strain curves predicted by the CG simulation-informed constitutive model is compared with experimental data.

\begin{figure}[h!]
	\centering
	\begin{subfigure}[b]{0.48\linewidth}
		\centering
		\includegraphics[width=\textwidth]{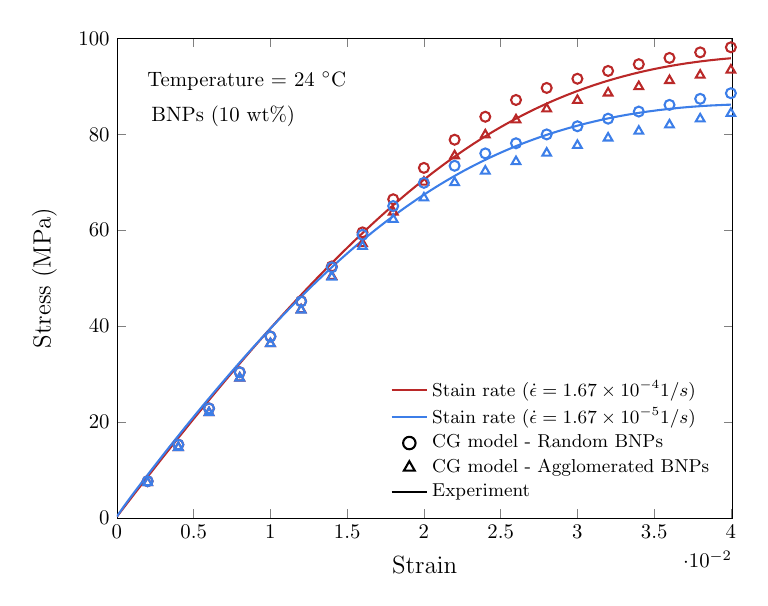}
		\caption{}
		\label{visco_NP01}
	\end{subfigure}
	\begin{subfigure}[b]{0.48\linewidth}
		\centering
		\includegraphics[width=\textwidth]{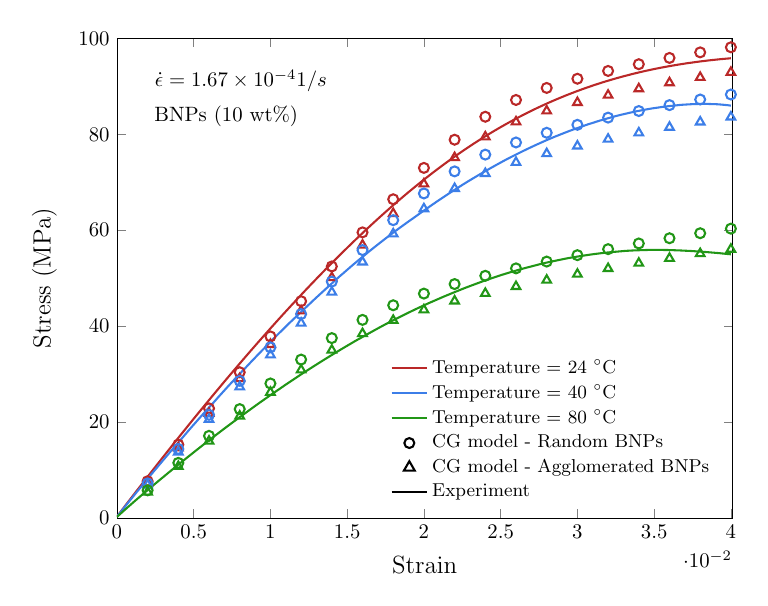}
		\caption{}
		\label{visco_NP02}  
	\end{subfigure}    
	\caption{Effect of strain rate (a) and temperature (b) on the stress–strain relationship of BNP(10 wt\%)/epoxy nanocomposites.} 
	\label{visco_NP0}
\end{figure}

Figs.~\ref{visco_NP01} show the stress–strain relationship of BNP(10 wt\%)/epoxy nanocomposite at the strain rate of $\textstyle 1.67\times 10^{-4}$ and $\textstyle 1.67\times 10^{-5}$~1/s. Also, the stress–strain curves predicted by the constitutive model are compared with experimental data at 24, 40 and 80~$^{\circ}$C in Fig.~\ref{visco_NP02}. As can be seen in the figures, the experimental stress–strain responses fall between the lower and upper bounds obtained from the constitutive model. This observation can be interpreted in the way that there is a nonuniform distribution of single-particle and agglomerated BNPs in the epoxy matrix. The nonuniform distribution results in a stress–strain response between the two extreme cases that the simulation-informed constitutive model predicts. Moreover, as expected, models based on well-dispersed and agglomerated BNPs provide upper and lower bounds for the stress–strain relationship. Both the overestimation and underestimation are in good agreement with experimental data.

\begin{figure}[h!]
	\centering
	\begin{subfigure}[b]{0.48\linewidth}
		\centering
	\includegraphics[width=1\textwidth]{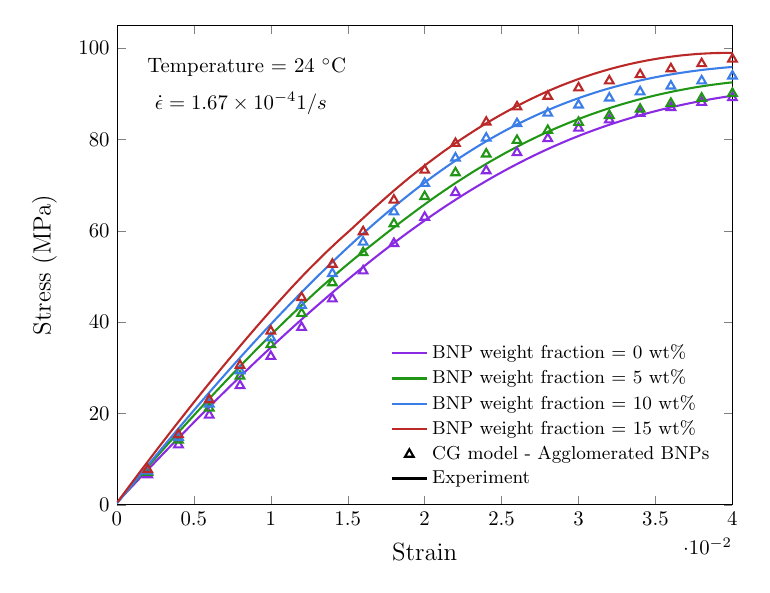}
		\caption{}
		\label{visco_NP_wt1}
	\end{subfigure}
	\begin{subfigure}[b]{0.48\linewidth}
		\centering
	\includegraphics[width=1\textwidth]{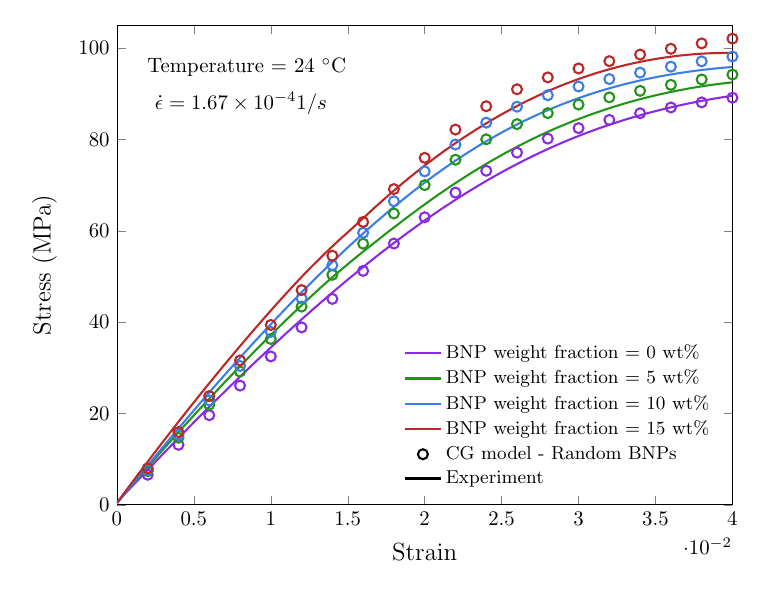}
		\caption{}
		\label{visco_NP_wt2}  
	\end{subfigure}    
	\caption{Effect of the epoxy matrix reinforced by (a) agglomerated BNPs and (b) well dispersed BNPs on the stress–strain relationship of BNP/epoxy nanocomposites at the room temperature.} 
	\label{visco_NP_wt}
\end{figure} 

The upper and lower bound predictions of BNP/epoxy nanocomposites' stress–strain response at different strain rates, temperatures, and BNP weight fractions are shown in Figs.~\ref{visco_NP_wt1} and ~\ref{visco_NP_wt2}, respectively. Comparing the predictions with experimental data allows a more comprehensive assessment of the predictive capabilities of the constitutive model. From Figs.~\ref{visco_NP_wt}, it can be seen that the linear elastic and nonlinear behavior of BNP/epoxy nanocomposites can be well predicted by the CG simulation-informed constitutive model. These results highlight the predictive capability of the constitutive model to reproduce the material behavior of BNP/epoxy nanocomposites across a range of strain rates, temperatures, and nanoparticle weight fractions.

\begin{table}[h!]
	\caption{Quantitative comparison between the constitutive model predictions 
		and experimental results in terms of stress error at $\varepsilon = 0.04$ 
		and root mean square error (RMSE) over the full stress--strain curve. 
		For nanocomposite systems, upper and lower bounds correspond to 
		well-dispersed (random) and fully agglomerated BNP models, respectively.}
	\label{tab:validation}
	\centering
	\begin{tabular}{
			>{\centering\arraybackslash}p{2.0cm}
			>{\centering\arraybackslash}p{1.2cm}
			>{\centering\arraybackslash}p{2.0cm}
			>{\centering\arraybackslash}p{2.8cm}
			>{\centering\arraybackslash}p{3.0cm}}
		\toprule
		{\normalsize Material} & 
		{\normalsize $T$ ($^{\circ}$C)} & 
		{\normalsize Strain rate (1/s)} & 
		{\normalsize $\sigma$ error -- upper/lower (\%)} & 
		{\normalsize RMSE -- upper/lower (MPa)} \\
		
		\cmidrule(r){1-5}
		
		Pure epoxy  & 24 & $1.67\times10^{-4}$ & 0.5         & 1.48 \\
		& 24 & $1.67\times10^{-5}$ & 1.7         & 1.32 \\
		& 40 & $1.67\times10^{-4}$ & 0.1         & 1.69 \\
		& 80 & $1.67\times10^{-4}$ & 6.5         & 1.22 \\
		\cmidrule(r){1-5}
		
		BNP 5~wt\%  & 24 & $1.67\times10^{-4}$ & 2.2~/~2.3  & 2.78~/~1.78 \\
		\cmidrule(r){1-5}
		
		BNP 10~wt\% & 24 & $1.67\times10^{-4}$ & 2.4~/~3.1  & 2.25~/~2.54 \\
		& 24 & $1.67\times10^{-5}$ & 2.8~/~2.1  & 1.44~/~2.84 \\
		& 40 & $1.67\times10^{-4}$ & 2.7~/~2.8  & 1.87~/~2.71 \\
		& 80 & $1.67\times10^{-4}$ & 9.7~/~1.9  & 2.28~/~1.98 \\
		\cmidrule(r){1-5}
		
		BNP 15~wt\% & 24 & $1.67\times10^{-4}$ & 3.2~/~1.3  & 2.48~/~2.90 \\
		\bottomrule
		
	\end{tabular}
\end{table}

To quantitatively assess the predictive capability of the constitutive model, Table 4 summarizes the relative error in stress at $\varepsilon = 0.04$ and the root mean square error (RMSE) over the full stress–strain curve for all tested configurations. The RMSE measures the average pointwise deviation between the predicted and experimentally measured stress across the entire strain range, providing a more comprehensive measure of agreement than a single-point comparison. For the pure epoxy system, the stress predictions remain within 1.7\% of the experimental values across both strain rates and all three temperatures, with RMSE values below 1.7 MPa in all cases — corresponding to less than 2\% of the peak stress. The largest deviation for the pure epoxy is observed at 80~$^{\circ}$C, where the stress error reaches 6.5\%. This is attributed to the combined influence of Kitagawa scaling sensitivity at larger offsets from reference and interpolation uncertainty in the CG-derived Argon model parameters near the upper calibration boundary. The relatively low RMSE of 1.22 MPa at this condition, despite the higher pointwise stress error of 6.5\%, reflects that the deviation is localized near the end of the strain range rather than distributed across the entire stress–strain curve.
For the BNP/epoxy nanocomposites at 24~$^{\circ}$C and 40~$^{\circ}$C, the stress predictions remain within approximately 3\% for both the well-dispersed and agglomerated bounds across all weight fractions and strain rates, with RMSE values consistently below 3 MPa. At 80~$^{\circ}$C, the agglomerated bound reproduces the experimental stress with an error of 1.9\%, while the well-dispersed bound shows a larger deviation of 9.7\%. This increased deviation at elevated temperature is consistent with the known reduction in model accuracy near the glass-transition region, as previously discussed above. Overall, the quantitative metrics confirm that the proposed CG simulation-informed constitutive model achieves a high level of predictive accuracy across a broad range of strain rates, temperatures, and nanoparticle weight fractions, while the upper and lower bounds consistently bracket the experimental data in a physically meaningful way.

\section{Conclusions}
\label{section:conclusions}

A molecular simulation-informed constitutive model that is able to capture the nonlinear viscoelastic damage behavior of BNP/epoxy nanocomposites is proposed. A combination of experimental data and simulation results was used to identify the parameters of the constitutive model. Thanks to large-scale CG simulations, the model can characterize the effect of BNP weight fraction and agglomerate size on the nanocomposites' stress–strain behavior. For this, a new amplification factor, accounting for the effect of BNP weight fraction and agglomerate size on the nanocomposites' nonlinear hyperelasticity and the linear response, was developed using CG simulations. Furthermore, the Argon viscoelastic model was calibrated using CG simulations to capture the effect of nanoparticles on the temperature-dependent viscoelastic flow in the nanocomposites. The modified Kitagawa model was also adopted to take into account the temperature dependency of the nonlinear hyperelasticity and the linear response.

The constitutive model is valid below the glass-transition region; however, prediction accuracy decreases at 80~$^{\circ}$C due to the increasing sensitivity of the Kitagawa exponential scaling at larger temperature offsets from reference and the interpolation uncertainty in the CG-derived Argon model parameters near the upper calibration boundary.
Furthermore, the validation is limited to uniaxial tensile loading at selected weight fractions, and the BNP distribution in the epoxy matrix is represented through upper and lower bounds rather than direct morphological characterization. Future work should address multiaxial loading conditions, hygrothermal effects, and a more explicit microstructure-informed description of nanoparticle dispersion.

Simulations of two extreme cases, including the epoxy with agglomerated and well-dispersed BNPs, reveal lower and upper bounds for the stress–strain response. Experimentally measured stress–strain curves fall between the numerically predicted upper and lower bounds. Experimental validation shows good agreement between the upper and lower bound predictions and the experimental data for a wide range of strain rates, temperatures, and BNP weight fractions. The experimental-numerical comparison confirms the capability of a simulation-based framework for calibrating predictive constitutive models for nanoparticle/epoxy nanocomposites. In addition, the framework significantly reduces the number of experimental tests required for developing constitutive models.


\bibliography{mybibfile}

\end{document}